\pdfoutput=1
\RequirePackage{ifpdf}
\ifpdf 
\documentclass[pdftex]{sigma}
\else
\documentclass{sigma}
\fi

\include{diagxy}
\input{xy}
\xyoption{all}

\newcommand{\Ss}{\textbf{s}}
\newcommand{\Tt}{\textbf{t}}

\newcommand{\im}{{\rm im}}

\renewcommand{\O}{\mathcal{O}}
\renewcommand{\S}{\mathcal{S}}

\renewcommand{\to}{\rightarrow}
\newcommand{\tto}{\rightrightarrows}

\DeclareMathOperator{\Ad}{Ad}

\DeclareMathOperator{\anch}{\mathbf{a}}
\DeclareMathOperator{\core}{\text{\upshape core}\,}

\newcommand{\gold}{\mscr{G}}

\usepackage{mathrsfs}
\newcommand{\mscr}[1]{\mathscr{#1}}

\newcommand{\tbp}{\nu} 

\newcommand{\sdp}{\ltimes}

\newcommand{\gpd}{\rightrightarrows}
\newcommand{\act}{\mathbin{\hbox{$<\kern-.4em\mapstochar\kern.4em$}}}

\newcommand{\ann}{0}

\newcommand{\Om}{\Omega}

\newcommand{\tilalpha}{\widetilde\Ss}
\newcommand{\tilbeta}{\skew6\widetilde\Tt}
\newcommand{\tillam}{\widetilde\lambda}
\newcommand{\tilone}{\widetilde 1}
\newcommand{\tilo}{\widetilde 0} 

\renewcommand{\Bar}[1]{\overline{#1}}
\newcommand{\barla}{\Bar{\lambda}}
\newcommand{\baralpha}{\Bar{\Ss}}
\newcommand{\barbeta}{\Bar{\Tt}}

\newcommand{\id}{\text{\upshape id}}

\newcommand{\Aut}{\operatorname{Aut}}

\newcommand{\R}{\mathbb{R}}

\numberwithin{equation}{section}

\newtheorem{Theorem}{Theorem}[section]
\newtheorem{Proposition}[Theorem]{Proposition}
 { \theoremstyle{definition}
\newtheorem{Definition}[Theorem]{Definition}
\newtheorem{Example}[Theorem]{Example}
\newtheorem{Remark}[Theorem]{Remark} }

\begin{document}

\allowdisplaybreaks

\newcommand{\arXivNumber}{1506.03216}

\renewcommand{\PaperNumber}{005}

\FirstPageHeading

\ShortArticleName{Poisson Geometry Related to Atiyah Sequences}

\ArticleName{Poisson Geometry Related to Atiyah Sequences}

\Author{Kirill MACKENZIE~$^\dag$, Anatol ODZIJEWICZ~$^\ddag$ and Aneta SLI{\.Z}EWSKA~$^\ddag$}

\AuthorNameForHeading{K.~Mackenzie, A.~Odzijewicz and A.~Sli{\.z}ewska}

\Address{$^\dag$~School of Mathematics and Statistics, University of Sheffield, Sheffield, S3 7RH, UK}
\EmailD{\href{mailto:K.Mackenzie@sheffield.ac.uk}{K.Mackenzie@sheffield.ac.uk}}

\Address{$^\ddag$~Institute of Mathematics, University in Bia{\l}ystok, Cio{\l}kowskiego 1M, 15-245 Bia{\l}ystok, Poland}
\EmailD{\href{mailto:aodzijew@uwb.edu.pl}{aodzijew@uwb.edu.pl}, \href{mailto:anetasl@uwb.edu.pl}{anetasl@uwb.edu.pl}} 

\ArticleDates{Received July 05, 2017, in f\/inal form January 06, 2018; Published online January 10, 2018}

\Abstract{We construct and investigate a short exact sequence of Poisson $\mathcal{V}\!\mathcal{B}$-groupoids which is canonically related to the Atiyah sequence of a~$G$-principal bundle~$P$. Our results include a description of the structure of the symplectic leaves of the Poisson groupoid $\frac{T^*P\times T^*P}{G}\tto \frac{T^*P}{G}$. The semidirect product case, which is important for applications in Hamiltonian mechanics, is also discussed.}

\Keywords{Atiyah sequence; $\mathcal{VB}$-groupoid; Poisson groupoid; dualization of $\mathcal{VB}$-groupoid}

\Classification{58H05; 22A22; 53D17}

\section{Introduction}

For many physical systems the conf\/iguration space is the total space of a principal bundle $P(M,G)$ over a base manifold $M$ which parametrizes the external degrees of freedom of the system under consideration. That is, there is a symmetry group $G$ which acts freely on $P$ with a quotient manifold $P/G = M$ and the projection $P\to M$ is locally trivial. In this situation the action of~$G$ lifts to the phase space~$T^*P$ with quotient manifold $T^*P/G$. The action of $G$ on~$T^*P$ is by symplectomorphisms and so the symplectic structure descends to a Poisson structure on~$T^*P/G$. This quotient space and its Poisson structure encode the mechanics and the symmetry of the original system.

The typical situation is when the Hamiltonian of the system is invariant with respect to the cotangent lift of the action of~$G$ on~$P$. Then the quotient manifold $T^*P/G$ becomes the reduced phase space of the system. For example the above happens in rigid body mechanics~\cite{MW} or when one formulates the equation of motion of a classical particle in a Yang--Mills f\/ield: the electromagnetic f\/ield is a particular case~\cite{Stern, Weinstein:1977}.

When the action of $G$ is lifted to $TP$ the quotient manifold $TP/G$ is the Atiyah algebroid of~$P(M.G)$, or equivalently the Lie algebroid of the associated gauge groupoid. There is a~natural structure of vector bundle on $TP/G$ with base $M$, and sections of this correspond to vector f\/ields on $P$ which are invariant under the group action; these are closed under the Lie bracket and def\/ine a bracket on the sections of $TP/G\to M$.

There are interesting geometric objects which include $TP/G$ and $T^*P/G$ as their crucial components. In this paper we will consider the Atiyah sequence (\ref{Aseq}) and the dual Atiyah sequence (\ref{Adual}) as well as the dual pair of Poisson manifolds (\ref{diagsymplpair}). From the dual Atiyah sequence we also construct a short exact sequence of $\mathcal{V}\!\mathcal{B}$-groupoids (\ref{duzyVtrojkaG}) over the gauge groupoid $\frac{P\times P}{G}\tto P/G$. The concept of $\mathcal{V}\!\mathcal{B}$-groupoid \cite{Pradines:1988} is recalled in Section \ref{sect:prelims}. The short exact sequences of $\mathcal{V}\!\mathcal{B}$-groupoids which we construct and study (see in particular the diagrams (\ref{duzyVtrojkaG}) and (\ref{C})) relate the various algebraic and Poisson structures inherent in the situation.

Even the case $P=G$ is very productive from the point of view of the applications in Hamiltonian mechanics as well as Lie groupoid theory. Note here that the cotangent groupoid $T^*G\tto T_e^*G$ is the symplectic groupoid of the Lie--Poisson structure of the dual $T^*_eG=T^*G/G$ of the Lie algebra $T_eG$ of $G$, see \cite{Wei}. As we will see, for a general principal bundle $P$ the cotangent groupoid $T^*G\tto T^*_e{G}$ is replaced by the symplectic groupoid $T^*\big(\frac{P\times P}{G}\big)\tto T^*P/G$ which one obtains by the dualization (in the sense of $\mathcal{V}\!\mathcal{B}$-groupoid dualization \cite{Mackenzie:GT, Pradines:1988}) of the tangent prolongation groupoid $T\big(\frac{P\times P}{G})\tto T(P/G\big)$ of the gauge groupoid.

We now describe the contents of each section in detail.

Section~\ref{sect:prelims} is concerned with aspects of $\mathcal{V}\!\mathcal{B}$-groupoids: their duality, principal actions of a Lie group $G$ on a~$\mathcal{V}\!\mathcal{B}$-groupoid, and short exact sequences of $\mathcal{V}\!\mathcal{B}$-groupoids. The main results are presented in Proposition~\ref{prop:sesvbgpds} and Proposition~\ref{prop:sesPBGVBgpds}, which show that for $\mathcal{V}\!\mathcal{B}$-groupoids, and for short exact sequences of $\mathcal{V}\!\mathcal{B}$-groupoids, the operations of dualization and quotient, commute.

In Section~\ref{sect:dAseq} we recall the def\/initions of Atiyah sequence and dual pair of Poisson manifolds and describe the relationship between these notions. Using this relationship we describe in detail the `double f\/ibration' structure of the symplectic leaves of $T^*P/G$; see Theorem~\ref{prop:lavessympl}.

In Section~\ref{sect:VBgauge}, we extend the Atiyah sequence (\ref{Aseq}) and its dual~(\ref{Adual}) to short exact sequences of $\mathcal{V}\!\mathcal{B}$-groupoids over the gauge groupoid, see (\ref{duzyVtrojkaG}) and (\ref{C}) respectively. From the point of view of Poisson geometry the most important structures are described by the diagram (\ref{C}), in which all objects are linear Poisson bundles and all arrows are their morphisms. Consequently, using~(\ref{C}), we obtain various relationships between the Poisson and vector bundle structures of $T^*P/G$ and $\frac{T^*P\times T^*P}{G}$, as well as the groupoid structures of $T^*\big(\frac{P\times P}{G}\big)\tto T^*P/G$ and $\frac{T^*P\times T^*P}{G}\tto T^*P/G$.

The case when $P$ is a group $H$ and $G=N$ is a normal subgroup of $H$ is discussed in Section~\ref{sect:eacr}. Some geometric constructions useful in Hamiltonian mechanics of semidirect products, see \cite{ratiu1, holm, ratiu2, ReymanSTS}, are also presented in this section.

\section[Short exact sequences of $\mathcal{V}\!\mathcal{B}$-groupoids]{Short exact sequences of $\boldsymbol{\mathcal{V}\!\mathcal{B}}$-groupoids}\label{sect:prelims}

The groupoids with which the paper will be chief\/ly concerned are the tangents and cotangents of gauge groupoids, and certain groupoids related to them. For any groupoid $\gold$ the tangent bundle $T\gold$ has both a Lie groupoid and a vector bundle structure and these are related in a~natural way. Owing to the relations between these structures, the cotangent bundle $T^*\gold$ also has a~natural Lie groupoid structure, and indeed is a symplectic groupoid with respect to the canonical cotangent symplectic structure \cite{CDW}.

The relations between the groupoid and vector bundle structures on $T\gold$ were abstracted by Pradines \cite{Pradines:1988} to the concept of $\mathcal{V}\!\mathcal{B}$-groupoid, and \cite{Pradines:1988} showed that for a general $\mathcal{V}\!\mathcal{B}$-groupoid $\Om$, dualizing the vector bundle structure on $\Om$ leads to a $\mathcal{V}\!\mathcal{B}$-groupoid structure on $\Om^*$. We recall this construction in Sections~\ref{subsect:VBgpds},~\ref{subsect:cores} and~\ref{subsect:duals}. For further detail, see \cite[Chapter~11]{Mackenzie:GT}.

In Section~\ref{sect:sesVBg} we show that, under a natural condition, short exact sequences of $\mathcal{V}\!\mathcal{B}$-groupoids may be dualized; see Proposition \ref{prop:sesvbgpds}. We then show that when the base of a $\mathcal{V}\!\mathcal{B}$-groupoid is the total space of a principal bundle, the group of which acts on, and preserves the structure of, the $\mathcal{V}\!\mathcal{B}$-groupoid, the quotient spaces form another $\mathcal{V}\!\mathcal{B}$-groupoid (Theorem~\ref{prop:pbgvb}). Finally in Section~\ref{subsect:quots} we show that the two processes~-- taking the quotient and taking the dual~-- commute (Proposition~\ref{prop:sesPBGVBgpds}).

\subsection[$\mathcal{V}\!\mathcal{B}$-groupoids]{$\boldsymbol{\mathcal{V}\!\mathcal{B}}$-groupoids} \label{subsect:VBgpds}

We consider manifolds with both a groupoid structure and a vector bundle structure. The compatibility condition between the groupoid and vector bundle structures can be described succinctly by saying that the groupoid multiplication must be a morphism of vector bundles and that the vector bundle addition must be a morphism of groupoids. In fact these conditions are expressed by a single equation. In order to formulate the equation several conditions on the source, target and bundle projections are necessary.

Consider a manifold $\Om$ with a groupoid structure on base $B$, and a vector bundle structure on base $\gold$. Here $B$ is to be a vector bundle on base $P$ and $\gold$ is to be a Lie groupoid on the same base $P$. (At present $P$ is an arbitrary manifold; later it will denote a principal bundle.) These structures are shown in Fig.~\ref{fig:VBg}.
\begin{figure}[h]
$$
\xymatrix@=1.4cm{
\Om \ar@<0.5ex>[d]^\tilalpha\ar@<-0.5ex>[d]_\tilbeta \ar[r]^\tillam &
\gold \ar@<-0.5ex>[d]_\Tt\ar@<0.5ex>[d]^\Ss \\
B \ar[r]^\lambda & P\\
}
$$
\caption{}\label{fig:VBg}
\end{figure}

The crucial equation, as described above, is
\begin{gather}\label{eq:icl}
(\xi_1 + \xi_2)(\eta_1 + \eta_2) = \xi_1\eta_1 + \xi_2\eta_2,
\end{gather}
where $\xi_i\in\Om$. In order for $\xi_1 + \xi_2$ to be def\/ined we must have $\tillam(\xi_1) = \tillam(\xi_2)$ where $\tillam\colon \Om\to \gold$ is the bundle projection. Likewise we must have $\tillam(\eta_1) = \tillam(\eta_2)$. For $\xi_1\eta_1$ to be def\/ined we must have $\tilalpha(\xi_1) = \tilbeta(\eta_1)$ and likewise we must have $\tilalpha(\xi_2) = \tilbeta(\eta_2)$. If we impose the conditions that the source and target projections $\Om\to B$ are morphisms of vector bundles, then it will follow that the product on the left hand side of (\ref{eq:icl}) is def\/ined. Likewise, if $\tillam$ is a~morphism of groupoids, then the addition on the right hand side will be def\/ined. We now state the formal def\/inition.

\begin{Definition}\label{defn:VBgpd}
A $\mathcal{V}\!\mathcal{B}$-groupoid consists of four manifolds $\Om$, $B$, $\gold$, $P$ together with Lie groupoid structures $\Om\gpd B$ and $\gold\gpd P$ and vector bundle structures $\Om\to\gold$ and $B\to P$, such that the maps def\/ining the groupoid structure (source, target, identity inclusion, multiplication) are morphisms of vector bundles, and such that the map
\begin{gather}\label{dblesource}
\xi \mapsto (\tilalpha(\xi), \tillam(\xi)),\qquad \Om\to B\times_P\gold = \{(b,\gamma)\,|\, \lambda(b) = \Ss(\gamma)\}
\end{gather}
formed from the bundle projection $\tillam$ and the source projection $\tilalpha$, is a surjective submersion.
\end{Definition}

The need for the f\/inal condition will be clear in Section~\ref{subsect:cores}. The map in (\ref{dblesource}) may be denoted $(\tilalpha,\tillam)$. Also, we refer to $\gold\gpd P$ as the \emph{side groupoid} and to $B\to P$ as the \emph{side vector bundle}.

The condition that the identity inclusion be a morphism of vector bundles is the f\/irst equation in (\ref{eq:sides}) below. It follows that $\tilone_{0_p} = \tilo_{1_p}$; that is, the groupoid identity over a zero element of $B$ is the zero element over the corresponding identity element of $\gold$. Further, $\tilone_{-b} = -\tilone_b$ for $b\in B$. Since multiplication and identity are morphisms of vector bundles, it follows in the usual way that inversion is also a morphism of vector bundles, and this is the second equation in~(\ref{eq:sides}),
\begin{gather}\label{eq:sides}
\tilone_{b_1+b_2} = \tilone_{b_1} + \tilone_{b_2},\qquad
(\xi + \eta)^{-1} = \xi^{-1} + \eta^{-1}.
\end{gather}
Next, the zero sections def\/ine a morphism of groupoids. To see this, f\/irst consider sources. By the def\/inition, $\tilalpha\circ\tilo = 0\circ\Ss$, since $\tilalpha$ and $\Ss$ constitute a morphism of vector bundles. But this can also be read as stating that $\tilo$ and $0$ commute with the sources. In the same way, $\tilo$ and $0$ commute with the targets and the multiplication, and are therefore morphisms of Lie groupoids. So we have the f\/irst two equations below
\begin{gather*}
\tilo_{\gamma_1\gamma_2} = \tilo_{\gamma_1} \tilo_{\gamma_2}, \qquad
\tilo_{\gamma^{-1}} = \tilo_\gamma^{-1},\qquad (-\eta)(-\xi) = -\eta\xi.
\end{gather*}
The third equation follows by a similar argument.

\subsection[The core of a $\mathcal{V}\!\mathcal{B}$-groupoid]{The core of a $\boldsymbol{\mathcal{V}\!\mathcal{B}}$-groupoid}\label{subsect:cores}

A necessary ingredient in the duality of $\mathcal{V}\!\mathcal{B}$-groupoids is the concept of core. The \emph{core} $K$ of a~$\mathcal{V}\!\mathcal{B}$-groupoid $(\Om;B;\gold;P)$ is the intersection of the kernel of the bundle projection $\tillam\colon \Om\to \gold$ with the kernel of the source $\tilalpha\colon \Om\to B$.

Equivalently, $K$ is the preimage under $(\tilalpha,\tillam)$ of the closed submanifold $\{(0_p,1_p)\,|\, p\in P\}$. Since $(\tilalpha,\tillam)$
is assumed to be a surjective submersion, $K$ is a closed submanifold of~$\Om$.

Equivalently again, $K$ is the pullback of $\ker (\tillam)$ along the unit map $P\to \gold$. This equips $K$ with a natural structure of vector bundle with base $P$.

\begin{Example}\label{ex:TG}
For any Lie groupoid $\gold\gpd P$, the groupoid $T\gold\gpd TP$ obtained by applying the tangent functor to the structure of $\gold$, is a $\mathcal{V}\!\mathcal{B}$-groupoid, $(T\gold;TP,\gold;P)$. The core is the Lie algebroid $A\gold$.

When $\gold$ is a Lie group $G$, the multiplication, denoted $\bullet$, in the tangent group $TG$ obtained by applying the tangent functor to the multiplication in~$G$, is given by the explicit formula
\begin{gather}\label{tangentG}
X_g \bullet Y_h= TL_g(h)Y_h+TR_h(g)X_g,
\end{gather}
where $L_g(h):=gh$ and $R_h(g):=gh$. In particular for $X_e, \ Y_e\in T_e G$ and zero elements $0_g\in T_gG$ and $0_h\in T_h G$ one has
\begin{gather}\label{tangentG1}
X_e\bullet Y_e=X_e+Y_e, \qquad 0_g\bullet 0_h= 0_{gh},\qquad 0_g\bullet X_e= TL_g(e)X_e.
\end{gather}
The group $TG$ is isomorphic to the semi-direct product $G\sdp T_eG$ with respect to the adjoint action.

For general groupoids $T\gold\gpd TP$ is not a groupoid semi-direct product.
\end{Example}

\subsection[The dual of a $\mathcal{V}\!\mathcal{B}$-groupoid]{The dual of a $\boldsymbol{\mathcal{V}\!\mathcal{B}}$-groupoid}\label{subsect:duals}

Since $\tillam\colon \Om\to\gold$ is a vector bundle, it has a dual bundle $\tillam_*\colon \Om^*\to\gold$. Somewhat surprisingly, this has a natural Lie groupoid structure with base $K^*$ and these structures make $(\Om^*;K^*,\gold;P)$ a $\mathcal{V}\!\mathcal{B}$-groupoid \cite{Pradines:1988}.

Take $\Phi\in\Om^*_\gamma$ where $\gamma = \tillam(\Phi)\in\gold$ has source $p = \Ss(\gamma)$ and target $q = \Tt(\gamma)$. Then the target and source of $\Phi$ in $K^*_q$ and $K^*_p$ respectively are def\/ined to be
\begin{gather*}
\big\langle\tilbeta_*(\Phi),k \big\rangle =
 \big\langle\Phi, k \tilo_\gamma\big\rangle, \quad k \in K_q,
\qquad
\langle\tilalpha_*(\Phi),k \rangle =
 \big\langle\Phi,-\tilo_\gamma k^{-1}\big\rangle, \quad k \in K_p.
\end{gather*}
The lack of symmetry is unavoidable: in def\/ining the core one must take either the kernel of $\Ss$ or of $\Tt$.

For the composition, take $\Psi\in\Om^*_\delta$ with $\tilalpha_*(\Psi) = \tilbeta_*(\Phi)$. To def\/ine $\Psi\Phi\in\Om^*_{\delta\gamma}$ we must def\/ine the pairing of $\Psi\Phi$ with any element of $\Om_{\delta\gamma}$. Now any element $\zeta$ of $\Om_{\delta\gamma}$ can be written as a~pro\-duct~$\eta\xi$ where $\eta\in\Om_\delta$ and $\xi\in\Om_\gamma$. We claim that
\begin{gather}\label{eq:dual2}
\langle\Psi\Phi, \eta\xi\rangle = \langle\Psi,\eta\rangle + \langle\Phi,\xi\rangle
\end{gather}
is well def\/ined. Any other choice of $\eta$ and $\xi$ must be $\eta\tau^{-1}$ and $\tau\xi$ for some $\tau$ with $\tilalpha(\tau) = \tilalpha(\eta) = \tilbeta(\xi)$. Also, $\tau$ must project to $1_q\in\gold$ where $q = \Ss(\delta)$, since $\eta\tau^{-1}$ must lie in the same f\/ibre over~$\gold$ as~$\eta$.

Write $b = \tilalpha(\tau)$. Then $\tau - \tilone_b$ is def\/ined and is a core element, say $k$. Now we have $\eta\tau^{-1} = \eta(\tilone_b+k)^{-1} =
(\eta + \tilo_\delta)(\tilone_b+k^{-1}) = \eta + \tilo_\delta k^{-1}$ so
\begin{gather*}
\big\langle\Psi, \eta\tau^{-1}\big\rangle = \big\langle\Psi, \eta\rangle + \langle\Psi, \tilo_\delta k^{-1}\big\rangle.
\end{gather*}
Proceeding in the same way, we likewise f\/ind that
\begin{gather*}
\langle\Phi, \tau\xi\rangle = \langle\Phi,\xi\rangle + \big\langle\Phi,k\tilo_\gamma\big\rangle.
\end{gather*}
From $\tilalpha_*(\Psi) = \tilbeta_*(\Phi)$ we know that $-\langle\Psi, \tilo_\delta k^{-1}\rangle = \langle\Phi,k\tilo_\gamma\rangle$, so (\ref{eq:dual2}) is well-def\/ined.

We now def\/ine the identity element of $\Om^*$ at $\chi\in K^*_p$. To do this we need to pair $\tilone_\chi$ with elements of $\Om$ which project to $1_p\in\gold$. Take such an element $\xi$ and write $b = \tillam(\xi)$. Then $\xi - \tilone_b$ is a core element $k$ and we def\/ine
\begin{gather*}
\big\langle\tilone_\chi,\tilone_b + k \big\rangle = \langle\chi,k \rangle.
\end{gather*}

It is now straightforward to check the proof of the following result.

\begin{Proposition} Given a $\mathcal{V}\!\mathcal{B}$-groupoid $(\Om, B, \gold; P)$, the construction above yields a~$\mathcal{V}\!\mathcal{B}$-groupoid $(\Om^*;K^*,\gold;P)$ with core $B^*$.
\end{Proposition}

Given $\omega\in B^*_p$, we identify it with the element $\overline{\omega}\in \Om^*$ def\/ined by
\begin{gather*}
\big\langle \overline{\omega}, \tilone_b + k\big\rangle = \big\langle \omega, b + \tilbeta(k)\big\rangle.
\end{gather*}

\begin{Example} Given a Lie groupoid $\gold\gpd P$, the dual of the $\mathcal{V}\!\mathcal{B}$-groupoid $(T\gold;\gold,TP;P)$ is $(T^*\gold;\gold,A^*\gold;P)$ where $A\gold$ is the Lie algebroid of~$\gold$. The groupoid $T^*\gold\gpd A^*\gold$ is a symplectic groupoid with respect to the canonical symplectic structure on $T^*\gold$ \cite{CDW,Weinstein:1987}.
\end{Example}

\subsection[Short exact sequences of $\mathcal{V}\!\mathcal{B}$-groupoids and their duals]{Short exact sequences of $\boldsymbol{\mathcal{V}\!\mathcal{B}}$-groupoids and their duals}\label{sect:sesVBg}
The main constructions of the paper rely on dualizing short exact sequences of $\mathcal{V}\!\mathcal{B}$-groupoids and in this subsection we give the basic results for this process. First we need the concept of morphism.

A \emph{morphism of $\mathcal{V}\!\mathcal{B}$-groupoids} $(\Om;B,\gold;P)\to (\Om';B',\gold';P')$ is a quadruple of maps $F\colon\Om\to\Om'$, $f_B\colon B\to B'$, $f_\gold\colon \gold\to\gold'$ and $f\colon P\to P'$ such that $(F, f_\gold)$ and $(f_B, f)$ are morphisms of vector bundles, and $(F, f_B)$ and $(f_\gold, f)$ are morphisms of Lie groupoids.

It follows that $F$ restricts to a vector bundle morphism of the cores $f_K\colon K\to K'$.

For the purposes of this paper we only need to consider the dualization of morphisms which preserve $\gold\gpd P$.

\begin{Proposition}\label{prop:dualF}
Let $F\colon \Om_1\to\Om_2$ and $f_B\colon B_1\to B_2$, together with the identity maps on $\gold$ and $P$, be a morphism of $\mathcal{V}\!\mathcal{B}$-groupoids $(\Om_1;B_1,\gold;P)\to (\Om_2;B_2,\gold;P)$. Then the dual maps $F^*\colon \Om_2^*\to \Om_1^*$ and $f^*_K\colon K_2^*\to K_1^*$ together with the identity maps on $\gold$ and $P$ are a morphism of $\mathcal{V}\!\mathcal{B}$-groupoids $(\Om_2^*;K_2^*,\gold;P)\to (\Om_1^*;K_1^*,\gold;P)$.
\end{Proposition}

The proof is a straightforward verif\/ication. We will apply the duality of $\mathcal{V}\!\mathcal{B}$-groupoids to structures in which the side groupoid is a
gauge groupoid, or is closely related to a gauge groupoid. At the moment we continue to allow the base manifold $P$ to be arbitrary; later in the section it
will denote a principal bundle.

A \emph{short exact sequence of $\mathcal{V}\!\mathcal{B}$-groupoids} consists of three $\mathcal{V}\!\mathcal{B}$-groupoids and two morphisms, as shown in Fig.~\ref{fig:sesVBg}, such that the three $\mathcal{V}\!\mathcal{B}$-groupoids have the same side groupoid $\gold\gpd P$ and the morphisms $(F, f_B)$ and $(H,h_B)$ preserve $\gold$ and $P$, and such that $\Om_1 \stackrel{F}{\to} \Om_2 \stackrel{H}{\to} \Om_3 $ is a short exact sequence of vector bundles over~$\gold$. This includes the conditions that the kernel of $F$ is the zero bundle over~$\gold$ and that the image of~$H$ is~$\Om_3$, as well as exactness at~$\Om_2$.
\begin{figure}[h]\centering
{\xymatrix@!
{
&&\Om_1 \ar[r] \ar@<-0.5ex>[d]\ar@<0.5ex>[d] \ar[rrd]_(.75){F} &
\gold \ar@<-0.5ex>[d]\ar@<0.5ex>[d] \ar@{=}[rrd] & &\\
&&B_1 \ar[r] \ar[rrd]_(.75){f_B} & P \ar@{=}[rrd] &
\Om_2 \ar[r] \ar@<-0.5ex>[d]\ar@<0.5ex>[d] \ar[rrd]_(.75){H} &
\gold \ar@<-0.5ex>[d]\ar@<0.5ex>[d] \ar@{=}[rrd] & &\\
&&&& B_2 \ar[r] \ar[rrd]_(.75){h_B} & P \ar@{=}[rrd]
& \Om_3 \ar[r] \ar@<-0.5ex>[d]\ar@<0.5ex>[d] &
\gold \ar@<-0.5ex>[d]\ar@<0.5ex>[d] \\
&&&&&& B_3 \ar[r] & P &
}}
\caption{} \label{fig:sesVBg}
\end{figure}

\begin{Proposition}\label{prop:sesvbgpds}
Consider a short exact sequence of $\mathcal{V}\!\mathcal{B}$-groupoids, as just defined and as shown in Fig.~{\rm \ref{fig:sesVBg}}. Then:
\begin{enumerate}\itemsep=0pt
\item[$(i)$] $ B_1 \stackrel{f_B}{\to} B_2 \stackrel{h_B}{\to} B_3$ is a short exact sequence of vector bundles over $P$;
\item[$(ii)$] $ K_1 \stackrel{f_K}{\to} K_2 \stackrel{h_K}{\to} K_3$ is a short exact sequence of vector bundles over $P$, where $K_i$ is the core of $\Om_i$, and $f_K$, $h_K$ are the induced morphisms of the cores;
\item[$(iii)$] the dualization of the short exact sequence of $\mathcal{V}\!\mathcal{B}$-groupoids over $\gold\gpd P$ results in the situation shown in Fig.~{\rm \ref{fig:DsesVBg}}, which is therefore also a short exact sequence of $\mathcal{V}\!\mathcal{B}$-groupoids.
\end{enumerate}
\end{Proposition}

\begin{proof} For (i) and (ii) there are six conditions to establish. We give three representative proofs.

First we prove that $f_B\colon B_1\to B_2$ is injective. Take $b_1\in B_1$ and suppose that $f_B(b_1) = 0$. Then $F(1_{b_1}) = 1_{f_B(b_1)} = 1_0$. It is also true that $F(1_0) = 1_{f_B(0)} = 1_0$ so since $F$ is injective we have $1_{b_1} = 1_0$ and therefore $b_1 = 0$.

Secondly we prove that $h_K\colon K_2\to K_3$ is surjective. Take $k_3\in K_3$. Since $K_3\subseteq\Om_3$ and $H\colon \Om_2\to\Om_3$ is surjective, there is $\xi\in\Om_2$ such that $H(\xi) = k_3$. Now if $\xi$ projects to $g\in\gold$ it follows that $k_3$ also projects to $g$; since $k_3$ is a core element we have $g = 1_m$ for some $m\in P$. Now write $X\in B_2$ for the source of~$\xi$. We have $h_B(X) = 0$.

The subtraction $\xi - \tilone_X$ is def\/ined since both project to $1_m$ in $\gold$. And $\xi - \tilone_X$ has source $X-X=0$ so is a core element. Now
$H(\xi - \tilone_X) = k_3-\tilone_0$ and $\tilone_0 = \tilo_1$, the zero element in the f\/ibre over $1_m$. So $\xi-\tilone_X$ is a core element which is mapped by~$H$ to $k_3$.

Thirdly we prove exactness at $K_2$. Since $f_K$ and $h_K$ are restrictions of $F$ and $H$ it follows from $H\circ F = 0$ that $h_K\circ f_K = 0$. Now suppose that $h_K(k_2) = 0$ where $k_2\in K_2$. By exactness at~$\Om_2$, there exists $\xi_1\in\Om_1$ such that $F(\xi_1) = k_2$. Write $b_1 = \tilalpha(\xi_1)$. Then $f_B(b_1) = 0$ and since~$f_B$ is injective, it follows that $b_1 = 0$. Now $\tillam(\xi_1) = \tillam(k_2) = 1_p$ for some $p\in P$, so $\xi_1\in K_1$. So the sequence of cores is exact at $K_2$.

The other conditions are proved in the same way.

Now apply Proposition \ref{prop:dualF} to $F$ and $H$ as morphisms of $\mathcal{V}\!\mathcal{B}$-groupoids. It follows that $F^*$ and $H^*$ are morphisms of $\mathcal{V}\!\mathcal{B}$-groupoids as shown in Fig.~\ref{fig:DsesVBg}.

Since $\Om_1 \stackrel{F}{\to} \Om_2 \stackrel{H}{\to} \Om_3$ is a~short exact sequence of vector bundles over $\gold$, the duals form a short exact sequence $\Om^*_3 \stackrel{H^*}{\to} \Om^*_2 \stackrel{F^*}{\to} \Om_1^*$, again of vector bundles over $\gold$. This completes the proof of~(iii).
\end{proof}

\begin{figure}[h]
\centering
{\xymatrix@!
{%
 && \Om_3^* \ar[r] \ar@<-0.5ex>[d]\ar@<0.5ex>[d] \ar[rrd]_(.75){H^*} &
\gold \ar@<-0.5ex>[d]\ar@<0.5ex>[d] \ar@{=}[rrd] & &\\
&& K_3^* \ar[r] \ar[rrd]_(.75){h_K^*} & P \ar@{=}[rrd] &
\Om_2^* \ar[r] \ar@<-0.5ex>[d]\ar@<0.5ex>[d] \ar[rrd]_(.75){F^*} &
\gold \ar@<-0.5ex>[d]\ar@<0.5ex>[d] \ar@{=}[rrd] & &\\
&&&& K^*_2 \ar[r] \ar[rrd]_(.75){f_K^*} & P \ar@{=}[rrd]
& \Om_1^* \ar[r] \ar@<-0.5ex>[d]\ar@<0.5ex>[d] &
\gold \ar@<-0.5ex>[d]\ar@<0.5ex>[d] \\
&&&&&& K_1^* \ar[r] & P & \\
}}
\caption{}\label{fig:DsesVBg}
\end{figure}

\subsection[Quotients of $\mathcal{V}\!\mathcal{B}$-groupoids over group actions]{Quotients of $\boldsymbol{\mathcal{V}\!\mathcal{B}}$-groupoids over group actions}\label{subsect:quots}

We now need to establish that this dualization process commutes with quotienting over a group action. In the cases which we consider, the base manifold $P$ is a principal $G$-bundle with projection $\mu\colon P\to P/G$ and we need to quotient over the action of $G$. In this situation the required quotient manifolds exist, and the constructions are straightfoward. Denote the group action by $\kappa\colon P\times G\to P$. Recall that the action in a principal bundle is free and that the orbits are equal to the f\/ibres of $\mu$. We always assume that the bundle is locally trivial.

First consider vector bundles over the total space $P$ of a principal bundle. A \emph{PBG-vector bundle over $P$} is a vector bundle $\lambda\colon E\to P$ together with an action $E\times G\to E$ by vector bundle automorphisms over the principal action $\kappa\colon P\times G\to P$.

Denote the orbit of $e\in E$ by $\langle e\rangle$ and the orbit of $p\in P$ by $\langle p\rangle$. The projection $\barla\colon E/G\to P/G$ is def\/ined by $\barla(\langle e\rangle) = \langle \lambda(e)\rangle$. Take $\langle e\rangle, \langle e'\rangle\in E/G$ such that $\barla(\langle e\rangle) = \barla(\langle e'\rangle)$. Then there exists a unique $g\in G$ such that $\lambda(e') = \lambda(e)g$. We def\/ine $\langle e\rangle + \langle e'\rangle = \langle eg + e'\rangle$ and $t\langle e\rangle = \langle te\rangle$ for $t\in\R$. It remains to prove that $E/G\to P/G$ is locally trivial; for this and the rest of the proof of the following proposition
see \cite[Section~3.1]{Mackenzie:GT}.

\begin{Proposition}\label{prop:pbgvb}
Let $\lambda\colon E\to P$ be a PBG-vector bundle over a principal bundle $P$ with group~$G$ and projection $\mu\colon P\to P/G$. Then the quotient manifold~$E/G$ exists, and has a vector bundle structure with base $P/G$, such that the natural projection $E\to E/G$ is a vector bundle morphism over~$\mu$.
\end{Proposition}

The case $E = TP$ arises in constructing the Atiyah sequence of a principal bundle.

There is a corresponding notion for Lie groupoids. A \emph{PBG-groupoid over $P(M,G)$} is a Lie groupoid $\gold\gpd P$ together with an action $\gold\times G\to \gold$ by groupoid automorphisms over the principal action $P\times G\to P$ \cite[ Def\/inition~2.5.4]{Mackenzie:GT}.

Denote the orbit of $\gamma\in \gold$ by $\langle \gamma\rangle$. The source and target maps $\baralpha\colon \gold/G\to P/G$ and $\barbeta\colon \gold/G\to P/G$ are def\/ined by $\baralpha(\langle \gamma\rangle) = \mu(\Ss(\gamma))$ and $\barbeta(\langle \gamma\rangle) = \mu(\Tt(\gamma))$. Given $\langle \gamma\rangle$ and $\langle \gamma'\rangle$ with $\baralpha(\langle \gamma\rangle) = \barbeta(\langle\gamma'\rangle)$ there exists a unique $g\in G$ such that $\Ss(\gamma)g = \Tt(\gamma')$. We def\/ine
\begin{gather*}
\langle \gamma\rangle\,\langle \gamma'\rangle = \langle (\gamma g)\gamma'\rangle.
\end{gather*}
That $\gold/G\gpd P/G$ is a groupoid is straightforward to check. The remaining details of the following proposition can be found in \cite[Section~3.1]{Mackenzie:GT}.
\begin{Proposition}\label{prop:pbggpd}
Let $\gold$ be a PBG-groupoid over a principal bundle $P$ with group $G$ and projection $\mu\colon P\to P/G$. Then the quotient manifold $\gold/G$ exists, and has a Lie groupoid structure with base $P/G$, such that the natural projection $\gold\to \gold/G$ is a morphism of Lie groupoids over~$\mu$.
\end{Proposition}

See \cite[Proposition~2.5.5]{Mackenzie:GT} for the proof. These two constructions can be combined so as to apply to $\mathcal{V}\!\mathcal{B}$-groupoids.

\begin{Definition}\label{defn:PBGVBgpd}
Let $P$ be a principal bundle with group $G$ and projection $\mu\colon P\to P/G$. A~\emph{PBG-$\mathcal{V}\!\mathcal{B}$-groupoid over} $P$ is a $\mathcal{V}\!\mathcal{B}$-groupoid $(\Om;B,\gold;P)$ together with right actions of $G$ on each of the manifolds $\Om$, $B$ and $\gold$ such that $\Om\gpd B$ and $\gold\gpd P$ are PBG-groupoids and $B\to P$ is a~PBG-vector bundle.
\end{Definition}

\begin{Proposition}\label{prop:pbgvbgpd}
Let $(\Om;B,\gold;P)$ be a PBG-$\mathcal{V}\!\mathcal{B}$-groupoid over $P$, described above, with co\-re~$K$. Then the quotient manifolds $\Om/G$, $B/G$, $\gold/G$ and $K/G$ exist, and form a $\mathcal{V}\!\mathcal{B}$-groupoid $(\Om/G;B/G,\gold/G;P/G)$ with core $K/G$ such that the natural maps
$\Om\to\Om/G$, $B\to B/G$, $\gold\to\gold/G$ and $P\to P/G$, constitute a~morphism of $\mathcal{V}\!\mathcal{B}$-groupoids.
\end{Proposition}

The proof only requires assembling the results of Propositions~\ref{prop:pbgvb} and~\ref{prop:pbggpd}.

We now need to establish that this quotienting process commutes with dualization. Consider a PBG-$\mathcal{V}\!\mathcal{B}$-groupoid $(\Om;B,\gold;P)$ and its dual $(\Om^*;K^*,\gold;P)$. Equip $\Om^*$ and $K^*$ with the contragredient actions of $G$; that is,
\begin{gather*}
\langle\Phi g,\xi\rangle = \big\langle \Phi,\xi g^{-1}\big\rangle,\qquad \langle\phi g,k\rangle = \big\langle \phi,k g^{-1}\big\rangle,
\end{gather*}
for $\Phi\in \Om_{\gamma}$, $\gamma\in\gold$, $\xi\in\Om_{\gamma g}$, $g\in G$, $\phi\in K^*$ and $k\in K$.

\begin{Proposition} With the structures just defined, $(\Om^*;K^*,\gold;P)$ is a PBG-$\mathcal{V}\!\mathcal{B}$-groupoid and the canonical maps
\begin{gather*}
\Om^*/G\to (\Om/G)^*, \qquad K^*/G\to (K/G)^*,
\end{gather*}
together with the identities on $\gold$ and $P$, constitute an isomorphism of $\mathcal{V}\!\mathcal{B}$-groupoids.
\end{Proposition}

The proof is a lengthy but straightforward verif\/ication.

Finally in this section we need to consider the preservation of exact sequences of PBG-$\mathcal{V}\!\mathcal{B}$-groupoids under dualization and quotient. The proof is a straightforward application of the techniques used above.

\begin{Proposition}\label{prop:sesPBGVBgpds}
In the short exact sequence shown in Fig.~{\rm \ref{fig:sesVBg}} assume that each $\mathcal{V}\!\mathcal{B}$-groupoid has the structure of a PBG-$\mathcal{V}\!\mathcal{B}$-groupoid with respect to a principal $G$-bundle structure on~$P$, and that $F$ and $H$ are $G$-equivariant. Equip the dual sequence, shown in Fig.~{\rm \ref{fig:DsesVBg}}, with the contragredient $G$-actions. Then the dual short exact sequence is also a short exact sequence of PBG-$\mathcal{V}\!\mathcal{B}$-groupoids and the canonical maps $\Om_i^*/G \to (\Om_i/G)^*$ and $K^*_i/G\to(K_i/G)^*$ together with the identities on~$\gold/G$ and $P/G$, constitute an isomorphism of $\mathcal{V}\!\mathcal{B}$-groupoids over $P/G$.
\end{Proposition}

\section{The dual Atiyah sequence}\label{sect:dAseq}

The Lie algebroid of a gauge groupoid is the Atiyah algebroid of the corresponding principal bundle~$P$, and this Lie algebroid is the central term of a short exact sequence of Lie algebroids, the Atiyah sequence. Accordingly, the dual of this Lie algebroid is the central term of a short exact sequence of vector bundles with Poisson structures. In this section we set up the notation needed for the study of this dual sequence. We also investigate the f\/ibre structure of the symplectic leaves of the Poisson manifold $T^*P/G$ using the notion of dual pair together with that of Atiyah sequence. The main results are collected in Theorem~\ref{prop:lavessympl}. Throughout the section, $P$~denotes a principal bundle unless otherwise specif\/ied.

\subsection{Principal bundles}

We consider a principal $G$-bundle $P$ with the notation of Section~\ref{subsect:quots}. One may take the tangent of the right action $\kappa\colon P\times G\to P$ and obtain the action{\samepage
\begin{gather*}
T\kappa\colon \ TP\times TG\to TP
\end{gather*}
 of the tangent group $TG$ on the tangent bundle $TP$.}

Applying the tangent functor to the multiplication in $G$ we obtain the group structure on~$TG$ given in~(\ref{tangentG}), and denoted by~$\bullet$. We see from (\ref{tangentG1}) that the zero section $0\colon G\to TG$ of the tangent bundle~$TG$ is a group monomorphism and one has the decomposition{\samepage
\begin{gather*}
TG=G\bullet T_e G \cong G\sdp T_eG
\end{gather*}
of $TG$ as a semi-direct product of $G$ and the normal subgroup $T_e G\subseteq TG$.}

We will use the following notations:
\begin{gather*}
\kappa(p,g)=\kappa_p(g)=\kappa_g(p)=pg
\end{gather*}
where $\kappa_p\colon G\to P$ and $\kappa_g\colon P\to P$. Thus we have
\begin{gather*}
T\kappa_p(g)\colon \ T_gG\to T_{pg}P \qquad\text{and}\qquad
T\kappa_g(p)\colon \ T_pP\to T_{pg}P\end{gather*}
for $g\in G$. The action $T\kappa\colon TP\times TG\to TP$ of the tangent group $TG$ on the tangent bundle $TP$ can be expressed by
\begin{gather}\label{actionTG}
T\kappa(p,g)(v_p,X_g)=T\kappa_g(p)v_p+T\kappa_p(g)X_g.
\end{gather}

In the following two propositions we collect various equalities and bundle isomorphisms which will be useful in what follows. The proof of the f\/irst proposition is a direct calculation.
\begin{Proposition}
For $g,h\in G$ and $p\in P$,
\begin{gather*}
\kappa_g\circ\kappa_p=\kappa_p\circ R_g,\\
\kappa_g\circ \kappa_p=\kappa_{pg}\circ I_{g^{-1}},\\
\kappa_{pg}=\kappa_p\circ L_g,\\
T\kappa_{pg}(e)=T\kappa_g(p)\circ T\kappa_p(e) \circ \Ad_g,\\
T\kappa_g(p)^{-1}=T\kappa_{g^{-1}}(pg),\\
T\kappa_{gh}(p)=T\kappa_h(pg)\circ T\kappa_g(p),\\
T^*\kappa_{gh}(p)=T^*\kappa_h(pg)\circ T^*\kappa_g(p),
\end{gather*}
where $I_g=L_g\circ R_{g^{-1}}$, $\Ad_g=TI_g(e)$ and $T^*\kappa_g(p)\colon T^*_pP\to T^*_{pg}P$ is defined by
\begin{gather*}
T^*\kappa_g(p):=\big(T\kappa_g(p)^{-1}\big)^*.
\end{gather*}
\end{Proposition}

Write $T^VP$ for the vertical subbundle of $TP$; that is, $T^VP = \ker T\mu$, and $T^{V*}P$ for its dual.

\begin{Proposition}There are the following isomorphisms of vector bundles:
\begin{subequations}
\begin{gather}
\label{v1} TP/T_eG\cong TP/T^VP,\\
\label{v2} TP/TG\cong (TP/T_eG)/G,\\
\label{v3} T(P/G)\cong TP/TG,\\
\label{v4} T^VP/G\cong P\times_{\Ad G}T_eG,\\
\label{v5} T^{V*}P/G\cong P \times _{\Ad^*G}T^*_eG,
\end{gather}
\end{subequations}
where $\langle p,X\rangle\in P\times_{\Ad G}T_eG$ is defined by $\langle p,X\rangle:=\{(pg,\Ad _{g^{-1}}X)\colon \ g\in G\}$.
\end{Proposition}

\begin{proof}Using the vector space isomorphisms \begin{gather*}
T\kappa_p(e)\colon \ T_eG\to T^V_pP \end{gather*}
and the action of the subgroup $T_eG\subseteq TG$ on $TP$ def\/ined by
\begin{gather*}
T_pP\times T_eG\ni(v_p,X_e)\mapsto v_p+T\kappa_p(e)X_e \in T_pP
\end{gather*}
we obtain (\ref{v1}).

In order to prove (\ref{v2}) we observe that
\begin{gather}
T\kappa_g(p)(v_p+T\kappa_p(e)X_e)
=T\kappa_g(p)v_p+(T\kappa_g(p)\circ T\kappa_p(e))X_e\nonumber\\
\hphantom{T\kappa_g(p)(v_p+T\kappa_p(e)X_e)}{} =T\kappa_g(p)v_p+T(\kappa_g\circ \kappa_p)(e)X_e
=T\kappa_g(p)v_p+T(\kappa_p\circ R_g)(e)X_e\nonumber\\
\hphantom{T\kappa_g(p)(v_p+T\kappa_p(e)X_e)}{} =T\kappa_g(p)v_p+T\kappa_p(g)(TR_g(e)X_e).\label{25}
\end{gather}
Assuming $X_g\!=\!TR_g(e)X_e$ we f\/ind from (\ref{25}) that the double quotient vector bundle $(TP/T_eG)\!/\!G$ is isomorphic to the quotient bundle $TP/TG$.

Since the f\/ibres of $T\mu$ are orbits of the action (\ref{actionTG}) we obtain the vector bundle isomorphism mentioned in (\ref{v3}).

Let us def\/ine actions $\phi_g\colon TP\to TP$ and $\phi_g^*\colon T^*P\to T^*P$ of $g\in G$ as follows
\begin{gather}\label{Phi}
\phi_g(v)(pg):=T\kappa_g(p)v \qquad {\rm and}\qquad \phi^*_g(\varphi)(pg):=T^*\kappa_g(p)\varphi
\end{gather}
for $ v\in T_pP$, $\varphi\in T^*_pP$. For the vector bundle trivialization $I\colon P\times T_eG\stackrel{\sim}{\longrightarrow} T^V P$ def\/ined by
\begin{gather}\label{Lambda}
I_{}(p,X_e):=T\kappa_p(e)X_e
\end{gather}
one has
\begin{gather}\label{diagLambda}
\phi_g\circ I = I\circ (\kappa_g\times \Ad_{g^{-1}}),
\end{gather}
where $g\in G$. Thus the bundle isomorphism $I_{}\colon P\times T_eG\stackrel{\sim}{\longrightarrow}T^VP$ def\/ines the isomorphism
\begin{gather*}[I]\colon \ P\times_{\Ad_G}T_eG\cong(P\times T_eG)/G\stackrel{\sim}{\longrightarrow}T^VP/G\end{gather*}
of quotient vector bundles presented in (\ref{v4}). Dualizing (\ref{diagLambda}) we obtain
the isomor\-phism (\ref{v5}).
\end{proof}

\subsection{The dual pair of Poisson manifolds}
We recall that the canonical 1-form $\gamma$ on $T^*P$ is def\/ined by
\begin{gather}\label{canform}
\langle \gamma_\varphi,\xi_\varphi\rangle:=
\langle \varphi,T\pi^*(\varphi)\xi_\varphi\rangle,
\end{gather}
where $\varphi\in T^*P$ and $\xi_\varphi\in T_\varphi(T^*P)$. Here $\pi^*\colon T^*P\to P$ is the projection
of $T^*P$ on the base $P$. We note that
\begin{gather}\label{invgamma} T\pi^*\circ T\phi _g^*=\phi_g\circ T\pi^*\end{gather}
for any $g\in G$. From the def\/inition (\ref{canform}) and (\ref{invgamma}) one has
\begin{gather*}
\langle\gamma _{\phi^*_g(\varphi)},T\phi^*_g(\varphi)\xi_{\varphi}\rangle=
\langle{\phi^*_g(\varphi)},T\pi^*(\phi^*_g(\varphi))T\phi^*_g(\varphi)\xi_{\varphi}\rangle
= \langle{\phi^*_g(\varphi)},(\phi_g\circ T\pi^*(\varphi))\xi_{\varphi}\rangle\\
\hphantom{\langle\gamma _{\phi^*_g(\varphi)},T\phi^*_g(\varphi)\xi_{\varphi}\rangle}{} = \langle T^*\kappa_g(p)\varphi, T\kappa_g(p)\circ T\pi^*(\varphi)\xi_\varphi\rangle
= \langle\varphi,T\kappa_g(p)^{-1}\!\circ T\kappa_g(p)\circ T\pi^*(\varphi)\xi_\varphi\rangle\\
\hphantom{\langle\gamma _{\phi^*_g(\varphi)},T\phi^*_g(\varphi)\xi_{\varphi}\rangle}{} =\langle\varphi,T\pi^*(\varphi)(\xi_{\varphi})\rangle
=\langle \gamma_{\varphi}, \xi_{\varphi}\rangle,\hfil
\end{gather*}
which means that $\gamma$ is invariant with respect to the action $\phi^*_g\colon T^*P\to T^* P$ def\/ined in (\ref{Phi}). This action generates two maps:
\begin{gather*}
\pi_G^*\colon \ T^*P\to T^*P/G, \qquad\text{and}\qquad J\colon \ T^*P\to T^*_e G,
\end{gather*}
where $\pi_G^*$ is the projection to the quotient manifold and $J$ is the $G$-equivariant
\begin{gather*} \label{xx3} J_{}(\phi_g^*\varphi)=\Ad^*_{g^{-1}}J_{}(\varphi)\end{gather*}
momentum map def\/ined by
\begin{gather}\label{momentum} J(\varphi):=\varphi\circ T\kappa_p(e),
\end{gather}
for $\varphi\in T^*_pP$.

One has the canonical Lie--Poisson structure on $T^*_eG$ def\/ined by
\begin{gather}\label{nawias}
\pi_{\text{L-P}}(Tf, Tg)(\chi):=\langle \chi,[Tf(\chi), Tg(\chi)]\rangle,
\end{gather}
where $f,g\in C^\infty(T^*_eG)$ and $[\cdot,\cdot]$ is the Lie bracket of the Lie algebra $T_eG$.

Since the symplectic form $d\gamma$ is invariant with respect to the action $\phi_g^*\colon T^*P\to T^*P$ of the group $G$, def\/ined in (\ref{Phi}), the Poisson bracket $\{f,g\}$ of $G$-invariant functions $f,g\in C^\infty(T^*P)$ is also an $G$-invariant function. So, the quotient manifold $T^*P/G$ is a Poisson manifold, the Poisson structure of which is def\/ined by the quotienting of the structure on $T^*P$.

Remembering that the action of $G$ on $P$ is free, we conclude that for any $p\in P$ the map $J_{}\colon T^*_pP\to T^*_e G$ is surjective. Thus we see that the momentum map $J_{}\colon T^*P\to T^*_e G$ is a~surjective submersion. So, one has two surjective Poisson submersions:
\begin{gather}\label{diagsymplpair}\unitlength=5mm \begin{split}
& \begin{picture}(11,4.6)(-1.5,0)
 \put(4.5,4.5){\makebox(0,0){$T^*P$}}
 \put(0,-0.5){\makebox(0,0){$T^*P/G$}}
 \put(9,-0.5){\makebox(0,0){$ T^*_e G$}}
 \put(4,4.2){\vector(-1,-1){4}}
 \put(5,4.2){\vector(1,-1){4}}
 \put(1.2,2.5){\makebox(0,0){$\pi^*_G$}}
 \put(7.5,2.5){\makebox(0,0){$J$}}
 \end{picture}\end{split}
\end{gather}
from the symplectic manifold $T^*P$, such that the Poisson subalgebras $(\pi^*_G)^* (C^\infty (T^*P/G))$ and $J^* (C^\infty(T_e^*G))$ commute, see~\cite{Wei}.
Therefore, the diagram (\ref{diagsymplpair}) gives a dual pair in the sense of the def\/inition in \cite[Section~9.3]{Wei}.

Let us mention that $\pi^*_G$ is a complete Poisson map; that is, the pullback of every function on~$T^*P/G$ that has a complete Hamiltonian vector f\/ield, also has a~complete Hamiltonian vector f\/ield (see \cite[Proposition~6.6]{Wei}). The completness of the momentum map $J$ follows from its $G$-equivariance.
In the rest of the paper we will assume that~$G$ and~$P$ are connected manifolds. This implies the connectness of the f\/ibres of $\pi^*_G$ and $J_{}$. Taking into consideration this assumption and the properties of the dual pair (\ref{diagsymplpair}) mentioned above one f\/inds (see \cite{Wei}) that there is a~one-to-one correspondence between the symplectic leaves
\begin{gather}\label{orbits}\S=\pi^*_G\big(J_{}^{-1}(\O)\big)\end{gather}
of $T^*P/G$ and the coadjoint orbits $ \O=J_{}((\pi^*_G)^{-1}(\S))$ which are the symplectic leaves of~$ T^*_e G$.

\subsection{The dual Atiyah sequence}
Using the vector bundle monomorphism $I\colon P\times T_eG\to T^V P\subset TP$, def\/ined in~(\ref{Lambda}), we obtain the following exact sequence
\begin{gather}\label{exact}
 P\times T_eG\stackrel{\ I}{\longrightarrow} TP\stackrel{A }{\longrightarrow} TP/T_eG\end{gather}
of vector bundles over $P$. The map $I$ is given by (\ref{Lambda}) and $A$ is a quotient map def\/ined as{\samepage
\begin{gather*}
A(v_p):=[v_p]=v_p+T\kappa_p(e)T_eG,\end{gather*}
where $v_p\in TP$.}

Quotienting (\ref{exact}) over the $G$-action and using the isomorphisms
(\ref{v2}), (\ref{v3}) and (\ref{v4}), we obtain the Atiyah sequence
\begin{gather}\label{Aseq}\begin{split}& \unitlength=5mm
\begin{picture}(11,5.6)(-4.5,0)
 \put(-2,5){\makebox(0,0){$P\times_{\Ad G}T_eG$}}
 \put(5,5){\makebox(0,0){$ TP/G$}}
 \put(12,5){\makebox(0,0){$T(P/G)$}}
 \put(-2,0){\makebox(0,0){$P/G$}}
 \put(5,0){\makebox(0,0){$P/G$}}
 \put(12,0){\makebox(0,0){$P/G$}}
 \put(5,4){\vector(0,-1){3}}
 \put(-2,4){\vector(0,-1){3}}
 \put(12.5,4){\vector(0,-1){3}}
 \put(7,5){\vector(1,0){2.7}}
 \put(6,0){\vector(1,0){4.7}}
 \put(-0.8,0){\vector(1,0){4.5}}
 \put(0.8,5){\vector(1,0){2.5}}
 \put(-1.5,2.5){\makebox(0,0){$\ $}}
\put(-3,2.5){\makebox(0,0){$[\operatorname{pr}_1]$}}
\put(13,2.5){\makebox(0,0){$\tilde\pi$}}
 \put(5.9,2.5){\makebox(0,0){$[\pi]$}}
 \put(12.5,2.5){\makebox(0,0){$\ $}}
 \put(8.5,5.4){\makebox(0,0){$\anch $}}
 \put(8.5,0.5){\makebox(0,0){$\id$}}
\put(1.5,0.5){\makebox(0,0){$ \id $}}
\put(1.8,5.4){\makebox(0,0){$\iota $}}
 \end{picture}\end{split}\end{gather}
where $\iota_{}=:[I_{}]$ and $\anch:=[A]$ is the anchor map, e.g., see~\cite{Mackenzie:GT}.

It follows from $\mu\circ \kappa_g=\mu$ that $T\mu\colon TP\to T(P/G)$ is constant on the orbits of the action $\phi_g\colon TP\to TP$, $g\in G$. Note that all terms of the above short exact sequence have a Lie algebroid structure over $P/G$ and the central term is the Atiyah algebroid. So, it follows from Lie algebroid theory (e.g., see \cite[Chapter~3]{Mackenzie:GT}) that the short exact sequence
\begin{gather}\label{Adual}\begin{split}& \unitlength=5mm
\begin{picture}(11,5.6)(-4.5,0)
 \put(-2,5){\makebox(0,0){$T^* (P/G)$}}
 \put(5,5){\makebox(0,0){$ T^*P/G$}}
 \put(12,5){\makebox(0,0){$P\times_{\Ad_G^*}T^*_eG$}}
 \put(-2,0){\makebox(0,0){$P/G$}}
 \put(5,0){\makebox(0,0){$P/G$}}
 \put(12,0){\makebox(0,0){$P/G$}}
 \put(5,4){\vector(0,-1){3}}
 \put(-2,4){\vector(0,-1){3}}
 \put(12,4){\vector(0,-1){3}}
 \put(7,5){\vector(1,0){2.7}}
 \put(6,0){\vector(1,0){4.7}}
 \put(-0.8,0){\vector(1,0){4.5}}
 \put(0,5){\vector(1,0){2.5}}
 \put(-2.5,2.5){\makebox(0,0){$\tilde\pi^*$}}
 \put(5.9,2.5){\makebox(0,0){$[\pi^*]$}}
 \put(12.9,2.5){\makebox(0,0){$[\operatorname{pr}_1] $}}
 \put(8.5,5.4){\makebox(0,0){$\iota^* $}}
 \put(8.5,0.5){\makebox(0,0){$\id $}}
\put(1.5,0.5){\makebox(0,0){$ \id$}}
\put(1.5,5.4){\makebox(0,0){$\anch^* $}}
 \end{picture}\end{split}\end{gather}
dual to the Atiyah sequence (\ref{Aseq}), is a short exact sequence of Poisson maps of linear Poisson bundles. Let us explain more precisely the above statement.

The action $\phi_g^*\colon T^*P\to T^*P,\ g\in G$, preserves the vector bundle structure $\pi^*\colon T^*P\to P$ of~$T^*P$ as well as its Poisson structure. Hence the quotient manifold $T^*P/G$ is a vector bundle $[\pi^*]\colon T^*P/G\to P/G$ over $P/G$. The linearity of the Poisson bracket $\{\cdot,\cdot\}$ on $C^\infty(T^*P/G)$ means that if $f,g\in C^\infty(T^*P/G)$ are linear on the f\/ibres of $[\pi^*]\colon T^*P/G\to P/G$ then the Poisson bracket $\{f,g\}$ has the same property. So, we have consistency between vector bundle and Poisson manifold structures of~$T^*P/G$. The above is also valid for the Poisson structure of~$T^*(P/G)$.

The linear Poisson structure of the bundle $P\times_{\Ad_G^*} T_e^*G\to P/G$ is def\/ined as follows. Let~$\pi_0$ be the zero Poisson tensor on $P$ and let $\pi_{L-P}$ be the Lie Poisson tensor on $T_e^*G$ def\/ined by~(\ref{nawias}). The product Poisson structure $\pi_0\times \pi_{L-P}$ def\/ined on $P\times T^*_eG$ is invariant with respect to the action $\phi_g\times \Ad_g^*\colon P\times T^*_eG\to P\times T^*_eG$, $g\in G$. Hence one has on the associated vector bundle $P\times_{\Ad_G^*} T_e^*G\to P/G$ the quotient linear Poisson structure. This property follows from the linearity and the $\Ad^*_G$-invariance of the Lie--Poisson bracket def\/ined in~(\ref{nawias}).

To conclude we mention that $\anch^*$ and $\iota^*$ are Poisson maps and preserve the vector bundle structures in the short exact sequence~(\ref{Adual}).

\subsection{Relationship between the dual pair and the dual Atiyah sequence}
As we have shown in the two previous subsections the principal $G$-bundle structure $\mu\colon P\to P/G$ of $P$ leads to two crucial, from the point of view of Poisson geometry, structures described in diagrams (\ref{diagsymplpair}) and (\ref{Adual}), respectively. We now discuss the relationship between them. For this reason, using the isomorphism $I\colon P\times T_eG\to T^VP $ def\/ined in (\ref{Lambda}), we consider the short exact sequence of vector bundles
\begin{gather*}
0\to T^{V0}P\stackrel{A^*}{\hookrightarrow} T^*P\stackrel{I^*}{\longrightarrow}P\times T_e^*G\to 0,\end{gather*}
dual to (\ref{exact}), where the subbundle $T^{V\ann}P\subseteq T^* P$ is the annihilator of $T^VP$ in $TP$;
that is, $T_p^{V\ann}P$ consists of those $\varphi\in T^*_p P$ which satisfy
\begin{gather}\label{TV0}
\varphi\circ T\kappa_p(e)=0.
\end{gather}
The vector bundle epimorphism $I^*$ dual to $I$ is related to the momentum map (\ref{momentum}) by
\begin{gather*}
I^*(\varphi)=(\pi^*(\varphi),J(\varphi));\end{gather*}
that is, $I^*=\pi^*\times J$ and the map $\iota^*\colon T^*P/G\to P\times _{\Ad_G^*}T_e^*G$ given by
\begin{gather*}
\iota^*=[I^*],\end{gather*}
is the quotient of $I^*$ over the group $G$. We see from (\ref{momentum}) and (\ref{TV0}) that $T^{V0}P=J^{-1}(0)$ which implies that $T^{V0}P$ is a~$G$-invariant vector subbundle of $T^*P$.

It follows from the Marsden--Weinstein symplectic reduction procedure \cite{MW}, that $T^{V0}P/G=J^{-1}(0)/G$ is a symplectic leaf of $T^*P/G$ corresponding to the one-element $\Ad_G^*$-orbit consisting of the zero element of $T^*_eG$. Proposition~\ref{thmquo} below shows that the dual anchor map $\anch^*\colon T^*(P/G)\to T^*P/G$ def\/ines a f\/ibrewise linear symplectic dif\/feomorphism between $T^*(P/G)$ and $J^{-1}(0)/G$. The result can also be found in \cite[Lemma~5.4]{ReymanSTS}. Note that this fact follows also from the general theory of Lie algebroids, where the dual $\anch^*$ of the anchor map~$\anch$ is a Poisson map from the symplectic manifold $T^*(P/G)$ to the symplectic leaf $J^{-1}(0)/G\subset T^*P/G$ which in this case is equal to $(\iota^*)^{-1}(P\times_{Ad^*_G} \{0\}) = T^{V0}P/G$. However it is interesting to prove this result explicitly.

 \begin{Proposition}\label{thmquo}
One has the vector bundle isomorphism
\begin{gather*} \unitlength=5mm\begin{picture}(11,4.6)(-1.5,0)
 \put(1,4.5){\makebox(0,0){$T^*(P/G)$}}
 \put(8,4.5){\makebox(0,0){$T^{V0}P/G$}}
 \put(1,-0.5){\makebox(0,0){$P/G$}}
 \put(8,-0.5){\makebox(0,0){$P/G$}}
 \put(1.2,3.5){\vector(0,-1){3}}
 \put(8.2,3.5){\vector(0,-1){3}}
 \put(3,4.5){\vector(1,0){3}}
 \put(2.4,-0.5){\vector(1,0){3.7}}
 \put(0.1,1.9){\makebox(0,0){$ \ $}}
 \put(0.5,1.9){\makebox(0,0){$ \tilde\pi $}}
 \put(9.1,1.9){\makebox(0,0){$[\pi^{*}]$}}
 \put(6.8,1.9){\makebox(0,0){$\ $}}
 \put(4.5,5.0){\makebox(0,0){$\anch^* $}}
 \put(4.5,0){\makebox(0,0){$ \id $}}
 \end{picture}\end{gather*}
 which is also a symplectomorphism.
 \end{Proposition}

\begin{proof}In order to describe $\anch^*$ in an explicit way we recall that $\mu\circ\kappa_g=\mu$ and thus
\begin{gather}\label{map1}
T\mu(pg)\circ T\kappa_g(p)=T\mu(p).
\end{gather}
Dualizing (\ref{map1}) and noting that $T\mu(p)^*\colon T_{\mu(p)}^*(P/G)\to T^{V0}_p P$ we obtain
\begin{gather}\label{mapdual}\begin{split}& \unitlength=5mm \begin{picture}(11,6)(-1.5,0)
 \put(1,3){\makebox(0,0){$T_{\mu(p)}^*(P/G)$}}
 \put(8,6){\makebox(0,0){$T^{V0}_p P$}}
 \put(8,0){\makebox(0,0){$T^{V0}_{pg} P$}}
 \put(2,4){\vector(2,1){4}}
 \put(8.2,1){\vector(0,1){4}}
 \put(2,2.5){\vector(2,-1){4}}
 \put(2.2,3.4){\makebox(0,0){$\ $}}
 \put(9.1,3.4){\makebox(0,0){$\ $}}
 \put(9.5,3.5){\makebox(0,0){$T\kappa_g^*(p)$}}
 \put(3.5,5.5){\makebox(0,0){$T\mu(p)^*$}}
 \put(3,1){\makebox(0,0){$T\mu(pg)^*$}}
\end{picture}\end{split}\end{gather}
where $T\mu(p)^*$, $T\mu(pg)^*$ are monomorphisms of vector spaces and $T^*\kappa_g(p)$ is a vector space isomorphism, for any $p\in P$ and $g\in G$.

From (\ref{mapdual}) we see that $\anch^*\colon T^*(P/G)\to T^{V0}P/G$ is given by
\begin{gather*}
T^*(P/G)\ni \rho\mapsto \anch^*(\rho)=\{T\mu(pg)^*\rho\colon \, g\in G\}\in T^{V0}P/G.
\end{gather*}

Now we will show that the canonical 1-form $\tilde\gamma$ of the cotangent bundle $T^*(P/G)$ is obtained as the reduction by $\anch^*$ of the canonical form $\gamma$ def\/ined in (\ref{canform}). For this reason we choose a~local trivialization
\begin{gather*}
\psi_\alpha\colon \ \mu^{-1}(\Omega_\alpha)\to \Omega_\alpha\times G,\end{gather*}
where $\cup_{\alpha\in I}\Omega_\alpha=M:=P/G$, of the principal bundle $P(M,G)$. Using local sections $s_\alpha\colon \Omega_\alpha\to \mu^{-1}(\Omega_\alpha)$, where $s_\alpha(m):=\psi^{-1}_\alpha(m,e)$, we def\/ine
\begin{gather*}
\anch^*_\alpha(\rho):=T\mu(s_\alpha(m))^*\rho
\end{gather*}
the maps $\anch^*_\alpha\colon (\tbp^*)^{-1}(\Omega_\alpha)\to (\mu\circ \pi^*)^{-1}(\Omega_\alpha)$ which ``trivialize '' the vector bundle isomorphism~$\anch^*$. Here $\nu\colon TM\to M$ and $\nu^*\colon T^*M\to M$ are the bundle projections.

Note that for $p=s_\alpha(m)$ one has
\begin{gather}
 \anch^*_\alpha(\rho)\circ T\kappa_p(e) =(T\mu(s_\alpha(m))^*\rho)\circ T\kappa_p(e)\nonumber\\
\hphantom{\anch^*_\alpha(\rho)\circ T\kappa_p(e)}{} =\rho\circ (T\mu(s_\alpha(m))\circ T\kappa_p(e))=\rho\circ T(\mu\circ\kappa_p)(e)=0.\label{d1}
\end{gather}
The last equality in (\ref{d1}) follows from the fact that the map $\mu\circ\kappa_p\colon G\to M$
\begin{gather*}(\mu\circ\kappa_p)(g)=\mu(pg)=\mu(p)\end{gather*}
is constant on $G$. Thus we f\/ind that $\anch^*_\alpha(\rho)\in T^{V0}P$.

Now for $\rho\in (\tbp^*)^{-1}(\Omega_\alpha)$ and $\varphi=\anch^*_\alpha(\rho)\in (\pi^*\circ \mu)^{-1}(\Omega_\alpha)$ we have
\begin{gather}
\langle((\anch^*_\alpha)^*\gamma_\rho,\xi_\rho \rangle =\langle \gamma_\varphi,T\anch^*_\alpha(\rho)\xi_\rho\rangle
= \langle \varphi, T\pi^*(\varphi)T\anch^*_\alpha(\rho)\xi_\rho\rangle = \langle \varphi, T(\pi^*\circ \anch^*_\alpha)(\rho)\xi_\rho\rangle\nonumber\\
\hphantom{\langle((\anch^*_\alpha)^*\gamma_\rho,\xi_\rho \rangle}{} = \langle T\mu(s_\alpha(\tbp^*(\rho))^*\rho, T(\tbp^*\circ \anch^*_\alpha)(\rho)\xi_\rho\rangle
 = \langle\rho, T\mu(s_\alpha(\tbp^*(\rho))T(\tbp^*\circ \anch^*_\alpha)(\rho)\xi_\rho \rangle\nonumber\\
\hphantom{\langle((\anch^*_\alpha)^*\gamma_\rho,\xi_\rho \rangle}{} = \langle \rho, T(\mu\circ\tbp^*\circ \anch^*_\alpha)(\rho)\xi_\rho \rangle = \langle \rho, T(\id)(\rho)\xi_\rho\rangle
= \langle\rho, \xi_\rho \rangle = \langle\tilde\gamma_\rho, \xi_\rho \rangle,\label{forma}
\end{gather}
where we used the following equalities
\begin{gather*}\mu\circ \pi^*\circ \anch^*_\alpha
=\id_{\Omega_\alpha} \qquad {\rm and} \qquad (s_\alpha\circ\tbp^*)(\rho)=(\pi^*\circ\anch^*_\alpha)(\rho).
\end{gather*}

From (\ref{forma}) we conclude that ${(\anch^*_\alpha)}^* \gamma=\tilde\gamma $ on $ (\tbp^*)^{-1}(\Omega_\alpha)$. The above shows that ${(\anch^*_\alpha)}^* \gamma={(\anch^*_\Tt)}^*\gamma$ on $\tilde \tbp^{-1}(\Omega_\alpha\cap\Omega_\Tt)$. As a consequence we obtain ${(\anch^*_\alpha)}^* \gamma=\tilde\gamma$.
\end{proof}

As we have seen, the dual pair (\ref{diagsymplpair}) gives a one-to-one correspondence between the coadjoint orbits $\O\subset T_e^*G$ of $G$ and the symplectic leaves $\S\subset T^*P/G$ of $T^*P/G$ given by (\ref{orbits}). Using the Atiyah dual sequence~(\ref{Adual}) we can investigate this correspondence in more detail. At f\/irst let us make the following observation:

 \begin{Proposition}\label{prop:orbits} For any coadjoint orbit $\O\subset T^*_eG$, we have
 \begin{gather}\label{orbits2} J_{}^{-1}(\O)/G={\iota_{}^*}^{-1}(P\times_{\Ad_G^*}\O).\end{gather}
 \end{Proposition}
\begin{proof} At f\/irst we note that
 \begin{gather}\label{orbits3} J_{}^{-1}(\O)=
\bigsqcup_{p\in P}\big(T_p^*P\cap J_{}^{-1}(\O)\big)={I_{}^*}^{-1}(P\times \O).\end{gather}
 Since $J_{}$ and $I_{}^*$ are $G$-equivariant maps and $\iota_{}^*=[I_{}^*]$ we
obtain (\ref{orbits2}) by quotienting (\ref{orbits3}) over~$G$.
\end{proof}

We observe that the vertical arrows in the diagram
\begin{gather}\label{part1} \begin{split}& \unitlength=5mm \begin{picture}(11,5.6)(-1.5,0)
 \put(1,5){\makebox(0,0){$T^*P/G$}}
 \put(8,5){\makebox(0,0){$P\times_{\Ad_G^*} T^*_eG$}}
 \put(1,0){\makebox(0,0){$P/G$}}
 \put(8,0){\makebox(0,0){$P/G$}}
 \put(1.2,4){\vector(0,-1){3}}
 \put(8.2,4){\vector(0,-1){3}}
 \put(3,5){\vector(1,0){2.6}}
 \put(2.4,0){\vector(1,0){3.7}}
 \put(0.5,2.4){\makebox(0,0){$[\pi^*]$}}
 \put(9.1,2.4){\makebox(0,0){$[\operatorname{pr}_1]$}}
 \put(4.5,5.5){\makebox(0,0){$\iota^*$}}
 \put(4.5,0.5){\makebox(0,0){$\id$}}
 \end{picture}\end{split}
\end{gather}
which is a part of the dual Atiyah sequence (\ref{Adual}), are the projections on the base of the corresponding vector bundle, whereas the map $\iota^*\colon T^*P/G\to P\times_{\Ad_G^*} T^*_eG$ is the bundle projection of an af\/f\/ine bundle over $P\times_{\Ad_G^*} T^*_eG$.

In order to see that the f\/ibre ${\iota^*}^{-1}(\langle p,\chi\rangle)$, where $\langle p,\chi\rangle \in [\operatorname{pr}_1]^{-1}(\langle p\rangle)$ and $(p,\chi)\in P\times T^*_eG$, is an af\/f\/ine space over the vector space $T^*_{\langle p\rangle}(P/G)$, let us take $\langle \varphi_1\rangle, \langle \varphi_2\rangle\in {\iota^*}^{-1}(\langle p,\chi\rangle )$. From $\im\anch^*=\ker\iota^*$ it follows that there exists a unique $\rho\in T^*_{\langle p\rangle}(P/G)$ such that
\begin{gather}\label{affine} \langle \varphi_1\rangle= \langle \varphi_2\rangle+\anch^*(\rho).\end{gather}
So, the vector space $T^*_{\langle p\rangle}(P/G)$ acts freely and transitively on ${\iota^*}^{-1}(\langle p,\chi\rangle)$. Note that
\begin{gather*} \dim {\iota^*}^{-1}(\langle p,\chi\rangle)=\dim T^*_{\langle p\rangle}(P/G).\end{gather*}

The af\/f\/ine f\/ibre bundle ${\iota^*}\colon T^*P/G\to P\times _{\Ad_G^*}T^*G$ can be described in the groupoid language. Namely, let us consider the cotangent bundle $T^*(P/G)$ as a groupoid $T^*(P/G)\tto P/G$ in which the source and target maps are equal to $\nu^*\colon T^*(P/G)\to P/G$. Then the dual anchor map $\anch^*\colon T^*(P/G)\to T^*P/G$ is the momentum map for the action of $T^*(P/G)\tto P/G$ on $T^*P/G$ def\/ined in (\ref{affine}). The orbits of this action are the f\/ibres of ${\iota^*}\colon T^*P/G\to P\times _{\Ad_G^*}T^*G$. From the above and from Proposition~\ref{prop:orbits} it follows that all
concerning (\ref{part1}) is also valid for the f\/ibration
\begin{gather}\label{part2}
\begin{split}& \unitlength=5mm\begin{picture}(11,5.6)(-1.5,0)
 \put(1,5){\makebox(0,0){$J^{-1}(\O)/G$}}
 \put(8,5){\makebox(0,0){$P\times_{\Ad_G^*}\O$}}
 \put(1,0){\makebox(0,0){$P/G$}}
 \put(8,0){\makebox(0,0){$P/G$}}
 \put(1.2,4){\vector(0,-1){3}}
 \put(8.2,4){\vector(0,-1){3}}
 \put(3,5){\vector(1,0){3}}
 \put(2.4,0){\vector(1,0){3.7}}
 \put(0.5,2.4){\makebox(0,0){$[\pi^*]$}}
 \put(9.1,2.4){\makebox(0,0){$[\operatorname{pr}_1]$}}
 \put(4.5,5.5){\makebox(0,0){$\iota^*$}}
 \put(4.5,0.5){\makebox(0,0){$\id$}}
 \end{picture}\end{split}\end{gather}

Let us mention that $T^*P/G$ does not have a f\/ibre structure over $T^*(P/G)$. However, choosing a section $\sigma\colon P\times _{\Ad_G^*}T^*_eG\to T^*P/G$ of the af\/f\/ine bundle ${\iota^*}\colon T^*P/G\to P\times _{\Ad_G^*}T^*_eG$ we obtain the vector bundle epimorphism $\tilde\sigma\colon T^*P/G\to T^*(P/G)$ def\/ined by the action (\ref{affine}) as follows
\begin{gather}\label{anchor}\anch^*(\tilde\sigma(\langle\varphi\rangle)
=\langle\varphi\rangle-\sigma(\langle p,\chi\rangle),\end{gather}
where $\iota^*(\langle \varphi\rangle)=\langle p,\chi\rangle$. Note here that $\anch^*$ is a monomorphism of vector bundles, thus the equali\-ty~(\ref{anchor}) def\/ines $\tilde\sigma(\langle\varphi\rangle)\in T^*(P/G)$ uniquely.

The following theorem summarizes the observations mentioned above .
 \begin{Theorem}\label{prop:lavessympl}\quad
\begin{enumerate}\itemsep=0pt
\item[$(i)$] The Poisson vector bundle $[\pi^*]\colon T^*P/G\to P/G$ has the structure of an affine bundle $\iota^*\colon T^*P/G\to P\times _{\Ad_G^*} T^*_eG$ over the total space of the vector bundle $[\operatorname{pr}_1]\colon P\times _{\Ad_G^*} T^*_eG\to P/G$, i.e., the fibre ${\iota^*}^{-1}(\langle p,\chi\rangle)$ of $\langle p,\chi\rangle\in P\times _{\Ad_G^*} T^*_eG$ is an affine space over the vector space $T^*_{\langle p\rangle}(P/G)$.
\item[$(ii)$] The symplectic leaf $\S=J^{-1}(\O)/G$ has the structure of an affine fibre bundle over the total space of the bundle $[\operatorname{pr}_1]\colon P\times _{\Ad_G^*} \O\to P/G$ with the orbit $\O\subset T^*_eG$ as a typical fibre.
\item[$(iii)$] Fixing a section $\sigma\colon P\times _{\Ad^*_G}\O\to T^*P/G$ we could consider the symplectic leaf $J^{-1}(\O)/G$ as a fibre bundle
 \begin{gather}\label{pisigma} \pi_\sigma\colon \ J^{-1}(\O)/G\to T^*(P/G)\end{gather}
 over the cotangent bundle $T^*(P/G)$ with $\O$ as the typical fibre. The total space and the base of {\rm (\ref{pisigma})} are symplectic manifolds. However, the bundle projection $\pi_\sigma$ is not a Poisson map in general.
\item[$(iv)$] In the case when $\O$ is a one-element orbit $\O=\{\chi\}$ (such an orbit corresponds to a~character of $G$) the bundle map $\pi_\sigma\colon J^{-1}(\chi)/G\to T^*(P/G)$ defines a diffeomorphism of manifolds, but it is not a symplectomorphism. The difference $\omega_\chi-\pi_\sigma^*d\tilde\gamma$ of symplectic forms, where $ \omega_\chi$ is the symplectic form of the symplectic leaf $J^{-1}(\chi)/G$ and $\tilde\gamma$ is the canonical symplectic form on $T^*(P/G)$, and is called the magnetic term, see {\rm \cite{Stern,Weinstein:1977}}.
\end{enumerate}
\end{Theorem}

 If $G$ is a commutative Lie group all the coadjoint orbits are one element sets $\O=\{\chi\}$, where $\chi\in T^*_eG$. Hence we have
\begin{gather*}
J^{-1}(\chi)/G ={\iota^*}^{-1}(P\times _{\Ad^*_G}\{\chi\})={\iota^*}^{-1}(P/G\times\{\chi\}) \nonumber\\
\hphantom{J^{-1}(\chi)/G}{} =\{\langle\varphi_p\rangle\in T^*P/G\colon \, \varphi_p\circ T\kappa_p(e)=\chi\},\nonumber
\end{gather*}
 where $\varphi_p\in T_p^*P$ and $\langle\varphi_p\rangle\in [\pi^*]^{-1}(\langle p \rangle)$. If $\varphi_p, \tilde\varphi_p\in T_p^*P$ satisfy
 \begin{gather*} \varphi_p\circ T\kappa_p(e)=\tilde\varphi_p\circ T\kappa_p(e)=\chi \end{gather*}
 then $\langle \varphi_p - \tilde\varphi_p\rangle \in J_{}^{-1}\{\O\}/G\cong T^*(P/G)$. Taking the above facts into account and identifying $P/G\times \{\chi\}$ with $P/G$ we conclude the following
 \begin{Remark} If $G$ is a commutative group then the symplectic leaves $J_{}^{-1}\{\chi\}/G$, $\chi\in T^*_eG$, are af\/f\/ine bundles over $P/G$ modeled over $J_{}^{-1}(0)/G\cong T^*(P/G)$, i.e., for $\langle p\rangle\in P/G$ the f\/ibres ${\iota_{}^*}^{-1}(\langle p\rangle)$ are af\/f\/ine spaces over the vector spaces $T^*_{\langle p\rangle}(P/G)$.
 \end{Remark}

A nice geometrical way to def\/ine a section $\sigma$ is given by the choice of a connection form $\alpha$ on the principal bundle $\mu\colon P\to P/G$ (see \cite{Weinstein:1977}), i.e., such $\alpha\in \Gamma^\infty(T^*P, T_eG)$ that
 \begin{gather}\label{alpha1} \alpha_p\circ T\kappa_p(e)=\id_{T_eG}\end{gather}
 and
 \begin{gather}\label{alpha2} \alpha_{pg}\circ T\kappa_g(p)=\Ad_{g^{-1}}\circ\alpha_p.\end{gather}
 Conditions (\ref{alpha1}) and (\ref{alpha2}) imply that the bundle epimorphism $\tilde\alpha\colon TP\to P\times T_eG$ def\/ined on~$TP$ by
 \begin{gather*}
 \tilde\alpha(v_p):=(p,\alpha_p(v_p)),\end{gather*}
 where $v_p\in T_pP$, after quotienting by $G$ gives the map
 \begin{gather}\label{alpha4} [\tilde\alpha]\colon \ TP/G\to P\times _{\Ad_G}T_eG.\end{gather}
 The dual of (\ref{alpha4})
 \begin{gather*}
 \sigma:=[\tilde\alpha]^*\colon \ P\times _{\Ad_G^*}T^*_eG\to T^*P/G\end{gather*}
 is the section of $\iota^*\colon T^*P/G\to P\times _{\Ad_G^*}T^*_eG$ mentioned above.

\section[Short exact sequences of $\mathcal{V}\!\mathcal{B}$-groupoids over gauge groupoids]{Short exact sequences of $\boldsymbol{\mathcal{V}\!\mathcal{B}}$-groupoids over gauge groupoids}\label{sect:VBgauge}

Applying the constructions investigated in Section~\ref{sect:prelims}, we will construct two short exact sequences, (\ref{duzyVtrojkaG}) and (\ref{C}), of $\mathcal{V}\!\mathcal{B}$-groupoids over the gauge groupoid $\frac{P\times P}{G}\tto P/G$. These are related to each other by the dualization procedure described in Section~\ref{sect:sesVBg}. Applying the results obtained in Section~\ref{sect:dAseq} we investigate the groupoid and Poisson structures of the objects involved~(\ref{C}).

\subsection[The tangent $\mathcal{V}\!\mathcal{B}$-groupoid of a gauge groupoid]{The tangent $\boldsymbol{\mathcal{V}\!\mathcal{B}}$-groupoid of a gauge groupoid}

 Let us consider the tangent $\mathcal{V}\!\mathcal{B}$-groupoid
\begin{gather*}
\begin{split}& \unitlength=5mm \begin{picture}(11,5.6)(-1.5,0)
 \put(1,5){\makebox(0,0){$T(P\times P)$}}
 \put(8,5){\makebox(0,0){$P\times P$}}
 \put(1,0){\makebox(0,0){$TP$}}
 \put(8,0){\makebox(0,0){$P$}}
 \put(1.2,4){\vector(0,-1){3}}
 \put(0.7,4){\vector(0,-1){3}}
 \put(8.2,4){\vector(0,-1){3}}
 \put(7.7,4){\vector(0,-1){3}}
 \put(3,5){\vector(1,0){3}}
 \put(2.4,0){\vector(1,0){3.7}}
 \put(2.2,2.4){\makebox(0,0){$T\operatorname{pr}_2$}}
 \put(9,2.4){\makebox(0,0){$\operatorname{pr}_2$}}
 \put(-0.3,2.4){\makebox(0,0){$T\operatorname{pr}_1$}}
 \put(7.1,2.4){\makebox(0,0){$\operatorname{pr}_1$}}
 \put(4.5,5.5){\makebox(0,0){$\ $}}
 \put(4.5,0.5){\makebox(0,0){$\ $}}
 \end{picture}\end{split}\end{gather*}
 of the pair groupoid $P\times P\tto P$. Since
 \begin{gather*} T\operatorname{pr}_1\times T\operatorname{pr}_2\colon \ T(P\times P)\tilde\to TP\times TP \end{gather*}
is a vector bundle isomorphism we will identify the tangent $\mathcal{V}\!\mathcal{B}$-groupoid $T(P\times P)\tto TP$ with the pair $\mathcal{V}\!\mathcal{B}$-groupoid $TP\times TP\tto TP$. By $T^V(P\times P)$ we denote the vertical subbundle of the tangent bundle $T(P\times P)$ def\/ined by the action
\begin{gather}\label{actionG}
\kappa_2\colon \ (P\times P)\times G\to(P\times P),\qquad ((p,q),g)\mapsto (pg,qg),
\end{gather}
of $G$ on the product $P\times P$. In consequence $X\in T_eG $ acts on $(v_p, w_q)\in TP\times TP$ as follows
\begin{gather*}
(v_p,w_q)\mapsto(v_p+T\kappa_p(e)X, w_q+T\kappa_q(e)X).
\end{gather*}
Note that $(v_p,w_q)=(T\operatorname{pr}_1(p,q)\times T\operatorname{pr}_2(p,q))v_{(p,q)}$, where $v_{(p,q)}\in T_{(p,q)}(P\times P)$, and thus
\begin{gather*}
T_{(p,q)}^V(P\times P):=\{(T\kappa_p(e)X,T\kappa_q(e)X)\in T_pP\times T_qP;\, X\in T_eG\}.
\end{gather*}
It follows from the properties of morphisms of $\mathcal{V}\!\mathcal{B}$-groupoids which preserve the side groupoids, that $T^V(P\times P)\tto T^VP$ is a subgroupoid of $T(P\times P)\tto TP$.

The groupoids mentioned above form a short exact sequence of $\mathcal{V}\!\mathcal{B}$-groupoids as def\/ined in Section~\ref{sect:sesVBg}
\begin{gather}\label{duzyV}\begin{split}& \unitlength=5mm \begin{picture}(11,5.6)(-4.5,0)
 \put(-2,5){\makebox(0,0){$T^V (P\times P)$}}
 \put(5,5){\makebox(0,0){$TP\times T P$}}
 \put(12,5){\makebox(0,0){$\frac{TP\times T P}{T_eG}$}}
 \put(-2,0){\makebox(0,0){$T^V P$}}
 \put(5,0){\makebox(0,0){$TP$}}
 \put(12,0){\makebox(0,0){${TP}/{T_eG}$}}
 \put(5.2,4){\vector(0,-1){3}}
 \put(4.7,4){\vector(0,-1){3}}
 \put(-2.3,4){\vector(0,-1){3}}
 \put(-1.8,4){\vector(0,-1){3}}
 \put(12.2,4){\vector(0,-1){3}}
 \put(11.7,4){\vector(0,-1){3}}
 \put(7,5){\vector(1,0){2.7}}
 \put(6,0){\vector(1,0){4}}
 \put(-0.8,0){\vector(1,0){4.5}}
 \put(0,5){\vector(1,0){2.5}}
 \put(8.5,5.4){\makebox(0,0){$A_2 $}}
 \put(8.5,0.5){\makebox(0,0){$A $}}
\put(1.5,0.5){\makebox(0,0){$ \ $}}
\put(1.5,5.4){\makebox(0,0){$\ $}}
 \end{picture}\end{split}\end{gather}
Here each side groupoid is the pair groupoid $P\times P\rightrightarrows P$, and $\frac{TP\times TP}{T_eG}\tto \frac{TP}{T_eG}$ is the quotient of the pair groupoid $TP\times TP\tto TP$ by $T_eG$.

 Now we def\/ine a $\mathcal{V}\!\mathcal{B}$-groupoid
\begin{gather}\label{trojkaTV} \begin{split}& \unitlength=5mm \begin{picture}(11,5.6)(-1.5,0)
 \put(1,5){\makebox(0,0){$P\times T_eG\times P$}}
 \put(8,5){\makebox(0,0){$P\times P$}}
 \put(0.7,0){\makebox(0,0){$P\times T_eG $}}
 \put(8,0){\makebox(0,0){$P$}}
 \put(1.2,4){\vector(0,-1){3}}
 \put(0.7,4){\vector(0,-1){3}}
 \put(8.2,4){\vector(0,-1){3}}
 \put(7.7,4){\vector(0,-1){3}}
 \put(3.3,5){\vector(1,0){2.7}}
 \put(2.7,0){\vector(1,0){3.5}}
 \put(2.2,2.4){\makebox(0,0){$\tilde\Ss$}}
 \put(9,2.4){\makebox(0,0){$\operatorname{pr}_2$}}
 \put(-0.3,2.4){\makebox(0,0){$\tilde\Tt$}}
 \put(7.2,2.4){\makebox(0,0){$\operatorname{pr}_1$}}
 \put(4.5,5.5){\makebox(0,0){$\ $}}
 \put(4.5,0.5){\makebox(0,0){$\ $}}
 \end{picture}\end{split}\end{gather}
 over the pair groupoid $P\times P\tto P$ as follows
\begin{gather*}
\begin{split} &\tilde\Ss(p,X,q):=(q,X),\qquad \tilde\Tt(p,X,q):=(p,X),\qquad \tilde\varepsilon(p,X):=(p,X,p),\\
& \tilde\iota(p,X,q):=(q,X,p),\qquad (p,X,q)(q,X,r):=(p,X,r),\end{split}
\end{gather*}
 where $p,q,r\in P$ and $X\in T_eG$, and the horizontal arrows in (\ref{trojkaTV}) are projections on the suitable terms of the product manifolds.

Let us def\/ine a map $I_2\colon P\times T_eG\times P\to T^V(P\times P)$ by
 \begin{gather*}
 I_2(p,X,q):=(T\kappa_p(e)X,T\kappa_q(e)X).\end{gather*}
 \begin{Proposition} The map $I_{}$ given by \eqref{Lambda}, and the map $I_2$ define an isomorphism
\begin{gather}\label{trivializTV} \begin{split}& \unitlength=5mm\begin{picture}(11,5.6)(-1.5,0)
 \put(1,5){\makebox(0,0){$P\times T_eG\times P$}}
 \put(8,5){\makebox(0,0){$T^V(P\times P)$}}
 \put(0.7,0){\makebox(0,0){$P\times T_eG $}}
 \put(8,0){\makebox(0,0){$T^VP$}}
 \put(1.2,4){\vector(0,-1){3}}
 \put(0.7,4){\vector(0,-1){3}}
 \put(8.2,4){\vector(0,-1){3}}
 \put(7.7,4){\vector(0,-1){3}}
 \put(3.5,5){\vector(1,0){2.5}}
 \put(2.4,0){\vector(1,0){3.7}}
 \put(2.2,2.4){\makebox(0,0){$\tilde\Ss$}}
 \put(9,2.4){\makebox(0,0){$\Ss$}}
 \put(-0.3,2.4){\makebox(0,0){$\tilde\Tt$}}
 \put(7.2,2.4){\makebox(0,0){$\Tt$}}
 \put(4.5,5.5){\makebox(0,0){$I_2 $}}
 \put(4.5,0.5){\makebox(0,0){$I_{} $}}
 \end{picture}\end{split}\end{gather}
 of $\mathcal{V}\!\mathcal{B}$-groupoids over $P\times P\tto P$.
\end{Proposition}
We conclude from the above that $I_{}$ and $I_2$ trivialize the vertical subbundles $T^VP\to P$ and $T^V(P\times P)\to P\times P$, respectively. Let us also note that the $\mathcal{V}\!\mathcal{B}$-groupoid isomorphism given in~(\ref{trivializTV}) is equivariant with respect to the action of the group $G$ def\/ined on $P\times T_eG\times P$ by
\begin{gather*} (p,X,q)\mapsto (pg, \Ad_{g^{-1}}X,qg)\end{gather*}
and on $T^V(P\times P)$ by
\begin{gather*}
(T\kappa_p(e)X, T\kappa_q(e)X)\mapsto
((T\kappa_g(p)\circ T\kappa_p(e))X,(T\kappa_g(q)\circ T\kappa_q(e))X)\\
\hphantom{(T\kappa_p(e)X, T\kappa_q(e)X)\mapsto}{} =((T\kappa_{pg}(e)\circ \Ad_{g^{-1}})X, (T\kappa_{qg}(e)\circ \Ad_{g^{-1}})X),
\end{gather*}
where $g\in G$. From (\ref{duzyV}) and the isomorphism (\ref{trivializTV}) we obtain the following short exact sequence of $\mathcal{V}\!\mathcal{B}$-groupoids
\begin{gather}\label{duzyVtrojka}\begin{split}& \unitlength=5mm\begin{picture}(11,5.6)(-4.5,0)
 \put(-2,5){\makebox(0,0){$P\times T_eG\times P$}}
 \put(5,5){\makebox(0,0){$TP\times T P$}}
 \put(12,5){\makebox(0,0){$\frac{TP\times T P}{T_eG}$}}
 \put(-2,0){\makebox(0,0){$P\times T_eG $}}
 \put(5,0){\makebox(0,0){$TP$}}
 \put(12,0){\makebox(0,0){${TP}/{T_eG}$}}
 \put(5.2,4){\vector(0,-1){3}}
 \put(4.7,4){\vector(0,-1){3}}
 \put(-2.3,4){\vector(0,-1){3}}
 \put(-1.8,4){\vector(0,-1){3}}
 \put(12.2,4){\vector(0,-1){3}}
 \put(11.7,4){\vector(0,-1){3}}
 \put(7,5){\vector(1,0){2.9}}
 \put(6,0){\vector(1,0){4.4}}
 \put(-0.4,0){\vector(1,0){4.4}}
 \put(0.5,5){\vector(1,0){2.5}}
 \put(8.5,5.4){\makebox(0,0){$A_2 $}}
 \put(8.5,0.5){\makebox(0,0){$A $}}
\put(1.5,0.5){\makebox(0,0){$ I_{}$}}
\put(1.5,5.4){\makebox(0,0){$ I_2$}}
 \end{picture}\end{split}\end{gather}
over $P\times P\tto P$ in which all arrows commute with the action of~$G$. Thus, after quotienting by~$G$, we obtain the short exact sequence of $\mathcal{V}\!\mathcal{B}$-groupoids
\begin{gather}\label{duzyVtrojkaG}\begin{split}& \unitlength=5mm\begin{picture}(11,8.6)(-7,0)
 \put(-5,8){\makebox(0,0){$\frac{P\times T_eG\times P}{G}$}}
 \put(-5,4){\makebox(0,0){$\frac{P\times T_eG}{G}$}}
 \put(-1,8){\makebox(0,0){$\frac{P\times P}{G}$}}
 \put(-1,4){\makebox(0,0){$ \frac{P}{G}$}}
 \put(-5.1,7.5){\vector(0,-1){2.7}}
 \put(-4.9,7.5){\vector(0,-1){2.7}}
 \put(-1.1,7.5){\vector(0,-1){2.7}}
 \put(-0.9,7.5){\vector(0,-1){2.7}}
 \put(-3,8){\vector(1,0){0.8}}
 \put(-4,4){\vector(1,0){2.2}}
 \put(-4.5,7.5){\vector(4,-1){4.4}}
 \put(-4.2,3.7){\vector(4,-1){5.2}}
 \put(-0,7.9){\line(4,-1){5}}
 \put(-0.1,7.8){\line(4,-1){5}}
 \put(-0.4,3.9){\line(4,-1){5.5}}
 \put(-0.5,3.8){\line(4,-1){5.5}}
 \put(2,6){\makebox(0,0){$\frac{TP\times TP}{G}$}}
 \put(2,2){\makebox(0,0){$TP/G$}}
 \put(6,6){\makebox(0,0){$\frac{P\times P}{G}$}}
 \put(6,2){\makebox(0,0){$ \frac{P}{G}$}}
 \put(1.9,5.5){\vector(0,-1){2.7}}
 \put(2.1,5.5){\vector(0,-1){2.7}}
 \put(5.9,5.5){\vector(0,-1){2.7}}
 \put(6.1,5.5){\vector(0,-1){2.7}}
 \put(4,6){\vector(1,0){0.8}}
 \put(3,2){\vector(1,0){2.2}}
 \put(2.5,5.5){\vector(4,-1){4.8}}
 \put(2.6,1.7){\vector(4,-1){4.7}}
 \put(6.4,2){\line(4,-1){5.9}}
 \put(6.3,1.9){\line(4,-1){5.9}}
 \put(7,6){\line(4,-1){5.5}}
 \put(6.9,5.9){\line(4,-1){5.5}}
 \put(9,4){\makebox(0,0){$T\big(\frac{P\times P}{G}\big)$}}
 \put(9,0){\makebox(0,0){$T(P/G)$}}
 \put(13,4){\makebox(0,0){$\frac{P\times P}{G}$}}
 \put(13,0){\makebox(0,0){$ \frac{P}{G}$}}
 \put(8.9,3.5){\vector(0,-1){2.7}}
 \put(9.1,3.5){\vector(0,-1){2.7}}
 \put(12.9,3.5){\vector(0,-1){2.7}}
 \put(13.1,3.5){\vector(0,-1){2.7}}
 \put(10.5,4){\vector(1,0){1.3}}
 \put(10.5,0){\vector(1,0){1.7}}
\end{picture}\end{split}\end{gather}
over the gauge groupoid $\frac{P\times P}{G}\tto {P}/{G}$. The oblique upper and lower arrows in (\ref{duzyVtrojkaG}) give the Atiyah sequences for the principal bundles $P\times P\to \frac{P\times P}{G}$ and $P\to { P}/{G}$, respectively. Note that the last groupoid in the short exact sequence~(\ref{duzyVtrojkaG}) is the tangent prolongation groupoid of the gauge groupoid $\frac{P\times P}{G}\tto P/G$.

\subsection[The cotangent $\mathcal{V}\!\mathcal{B}$-groupoid of a gauge groupoid]{The cotangent $\boldsymbol{\mathcal{V}\!\mathcal{B}}$-groupoid of a gauge groupoid}\label{sec:cotangauge}

Now let us dualize (\ref{duzyVtrojkaG}) in the sense of Section~\ref{sect:sesVBg}. Since dualization commutes with the action of $G$ we start from the dualization of (\ref{duzyVtrojka}).

\begin{Proposition}\label{prop18} There are the following vector bundle isomorphisms:
\begin{enumerate}\itemsep=0pt
\item[$(i)$] $\core( TP\times TP)\cong TP $,
\item[$(ii)$] $\core (T^V(P\times P))\cong \core (P\times T_eG\times P)\cong P\times\{0\}$,
\item[$(iii)$] $\core\big(\frac{TP\times TP}{T_eG}\big)\cong \ TP$.
\end{enumerate}
\end{Proposition}

\begin{proof}
(i) From the def\/inition we see that $(v_p,w_q)\in \core( TP\times TP)$ if and only if $p=q$ and $w_q=0$. Thus $\core( TP\times TP)=TP\times(\{0\}\times P)\cong TP$.

(ii) The element $(p,X,q)\in \core (P\times T_eG\times P)$ if and only if $p=q$ and $X=0$. Thus
$\core((P\times T_eG\times P))=\{0\}\times P$.

(iii) An element $\langle v_p,w_q\rangle\in \frac{TP\times T P}{T_eG}$ is def\/ined by
\begin{gather}\label{Xquot}
\langle v_p,w_q\rangle:=\{(v_p+T\kappa_p(e)X,\, w_q+T\kappa_q(e)X);\, X\in T_eG\}.
\end{gather}
So, $\langle v_p,w_q\rangle\in \core\big(\frac{TP\times T P}{T_eG}\big)$ if and only if $p=q$ and $w_q=T\kappa_q(e)Y$ for some $Y\in T_eG$. Choosing in~(\ref{Xquot}) $X=-Y$ we f\/ind that
\begin{gather*}
\core\big(\tfrac{TP\times T P}{T_eG}\big)=\{\langle v_p,0_p\rangle;\, v_p\in TP\}\cong TP.\tag*{\qed}
\end{gather*}\renewcommand{\qed}{}
\end{proof}

Applying the dualization procedure presented in Section~\ref{sect:sesVBg} and Proposition \ref{prop18} we see that the short exact $\mathcal{V}\!\mathcal{B}$-groupoid sequence (over $P\times P\tto P$) dual to (\ref{duzyV}) is
 \begin{gather}\label{duzyVdual}\begin{split}& \unitlength=5mm\begin{picture}(11,5.6)(-4.5,0)
 \put(-2,5){\makebox(0,0){$T^{V0}(P\times P)$}}
 \put(5,5){\makebox(0,0){$T^*P\times T ^*P$}}
 \put(12,5){\makebox(0,0){$P\times T^*_eG\times P$}}
 \put(-2,0){\makebox(0,0){$T^*P$}}
 \put(5,0){\makebox(0,0){$T^*P$}}
 \put(12.5,0){\makebox(0,0){$P\times\{0\}^*\cong P$}}
 \put(5.2,4){\vector(0,-1){3}}
 \put(4.7,4){\vector(0,-1){3}}
 \put(-2.3,4){\vector(0,-1){3}}
 \put(-1.8,4){\vector(0,-1){3}}
 \put(12.2,4){\vector(0,-1){3}}
 \put(11.7,4){\vector(0,-1){3}}
 \put(7,5){\vector(1,0){2.2}}
 \put(6,0){\vector(1,0){3.7}}
 \put(-0.8,0){\vector(1,0){4.5}}
 \put(0,5){\vector(1,0){2.5}}
 \put(1.5,5.4){\makebox(0,0){$A_2^*$}}
 \put(8.5,5.4){\makebox(0,0){$I_2^* $}}
 \put(8.5,0.5){\makebox(0,0){$\pi^* $}}
\put(1.5,0.5){\makebox(0,0){$\id $}}
\put(1.5,5.4){\makebox(0,0){$\ $}}
 \end{picture}\end{split}\end{gather}
where the dual $A_2^*$ of $A_2$ is the inclusion map and $I_2^*$ is given by
\begin{gather*} I_2^*(\varphi,\psi):=(p,\varphi_p\circ T\kappa_p(e)+\psi_q\circ T\kappa_q(e),q)\end{gather*}
for $p=\pi^*(\varphi)$ and $q=\pi^*(\psi)$. The map $I_2^*\colon T^*P\times T^*P\to P\times T^*_eG\times P$ is a groupoid morphism over the bundle projection $\pi^*\colon T^*P\to P$.

We note here that for $T^*P\times T^*P\tto T^*P$ we have
\begin{gather}
 \Ss(\varphi_p,\psi_q)=-\psi_q,\qquad
 \Tt(\varphi_p,\psi_q)=\varphi_p,\qquad
 \varepsilon(\varphi_p)=(\varphi_p,-\varphi_p),\nonumber\\
 \iota(\varphi_p,\psi_q)=(-\psi_q, -\varphi_p),\qquad
(\varphi_p,\psi_q)(-\psi_q,\lambda_r)=(\varphi_p,\lambda_r),\label{s}\end{gather}
 as the source map, target map, identity section, inverse map and groupoid product, respectively.

\begin{Remark} It is easy to see that the involution $\delta\colon T^*P\times T^*P\to T^*P\times T^*P$ def\/ined by
\begin{gather*} \delta(\varphi_p,\psi_q)=(\varphi_p,-\psi_q) \end{gather*}
gives an isomorphism of symplectic groupoids between the pair groupoid $T^*P\times T^*P\tto T^*P$ with the symplectic form $d(\operatorname{pr}_1^*\gamma-\operatorname{pr}_2^*\gamma)$ on $T^*P\times T^*P$ and the symplectic groupoid $T^*P\times T^*P\tto T^*P$ for which the groupoid structure is given in (\ref{s}) and the symplectic form is $d(\operatorname{pr}_1^*\gamma+\operatorname{pr}_2^*\gamma)$.
\end{Remark}

The groupoid maps for the groupoid $P\times T_e^*G\times P\tto P$ are def\/ined in the following way
\begin{gather*}
\Ss(p,\mathcal{X},q):=q,\qquad \Tt(p,\mathcal{X},q):=p,\qquad \varepsilon(p):=(p,0,p),\qquad
\iota(p,\mathcal{X},q):=(q,-\mathcal{X},p),
\end{gather*}
and the groupoid product of $(p,\mathcal{X},q),\ (q,\mathcal{Y},r)\in P\times T^*_eG\times P$ is def\/ined as
\begin{gather*}
(p,\mathcal{X},q)(q,\mathcal{Y},r):=(p,\mathcal{X}+\mathcal{Y},r).
\end{gather*}

The second component of the bundle epimorphism $I_2^*$ def\/ines the momentum map
\begin{gather*}
J_2(\varphi_p,\psi_q)=\varphi_p\circ T\kappa_p(e)+\psi_q\circ T\kappa_q(e)
\end{gather*}
for the cotangent bundle $T^*(P\times P)=T^*P\times T^*P$ whose symplectic structure is given by the canonical symplectic form $d\gamma_2=d(\operatorname{pr}_1^*\gamma+\operatorname{pr}_2^*\gamma)$ and the action of the group $G$ on $P\times P$ is def\/ined in~(\ref{actionG}).

In order to see that $J_2\colon T^*P\times T^*P\to T_e^*G$ is a momentum map we note that $(P\times P,\mu_2\colon P\times P\to \frac{P\times P}{G}, G)$ is a $G$-principal bundle and apply the same consideration as for $J\colon T^*P\to T^*_e G$ in~(\ref{momentum}).

We note also that the vector bundle $\big(\frac{TP\times T P}{T_eG}\big)^*\to P\times P$ is isomorphic to the vector subbundle $T^{V0}(P\times P)\to P\times P$ which contains such $(\varphi_p,\psi_q)\in T^*_pP\times T^*_q P$ that
\begin{gather*} \varphi_p\circ T\kappa_p(e)+\psi_q\circ T\kappa_q(e)=0.\end{gather*}
Hence we can identify the dual $\mathcal{V}\!\mathcal{B}$-groupoid $\big(\frac{TP\times T P}{T_eG}\big)^*\tto T^*P$ with the subgroupoid $T^{V0}(P\times P)\tto T^*P$ of the dual $\mathcal{V}\!\mathcal{B}$-groupoid $T^*P\times T^*P\tto T^*P$.

For any $g\in G$ one has
\begin{gather*} I_2^*(T^*\kappa_g(p)\varphi_p, T^*\kappa_g(q)\psi_q)=(pg, \Ad^*_{g^{-1}}(\varphi_p\circ T\kappa_g(e)+\psi_q\circ T\kappa_g(e)), qg), \end{gather*}
where $(\varphi_p,\psi_q)\in T^*_pP\times T^*_qP$. Thus all horizontal arrows in (\ref{duzyVdual}) def\/ine $G$-equivariant $\mathcal{V}\!\mathcal{B}$-groupoid morphism. All groupoid maps and products for the groupoids in (\ref{duzyVdual}) are also $G$-equivariant.

So, according to Proposition \ref{prop:sesPBGVBgpds}, quotienting (\ref{duzyVdual}) by $G$ we obtain the short exact sequence of $\mathcal{V}\!\mathcal{B}$-groupoids over the gauge groupoid $\frac{P\times P}{G}\tto {P}/{G}$
 \begin{gather}\label{C}\begin{split}& \unitlength=5mm\begin{picture}(11,8.6)(-7,0)
 \put(-5,8){\makebox(0,0){$T^*\big(\frac{P\times P}{G}\big)$}}
 \put(-5,4){\makebox(0,0){$T^*P/G$}}
 \put(-1,8){\makebox(0,0){$\frac{P\times P}{G}$}}
 \put(-1,4){\makebox(0,0){$ \frac{P}{G}$}}
 \put(-5.1,7.5){\vector(0,-1){2.7}}
 \put(-4.9,7.5){\vector(0,-1){2.7}}
 \put(-1.1,7.5){\vector(0,-1){2.7}}
 \put(-0.9,7.5){\vector(0,-1){2.7}}
 \put(-3,8){\vector(1,0){0.8}}
 \put(-4,4){\vector(1,0){2.2}}
 \put(-4.5,7.5){\vector(4,-1){4.4}}
 \put(-4.2,3.7){\vector(4,-1){5.2}}
 \put(-0,7.9){\line(4,-1){5}}
 \put(-0.1,7.8){\line(4,-1){5}}
 \put(-0.4,3.9){\line(4,-1){5.5}}
 \put(-0.5,3.8){\line(4,-1){5.5}}
 \put(-3,6.5){\makebox(0,0){$\anch_2^* $}}
 \put(-2,2.6){\makebox(0,0){$ \id$}}
 \put(2,6){\makebox(0,0){$\frac{T^*P\times T^*P}{G}$}}
 \put(2,2){\makebox(0,0){$T^*P/G$}}
 \put(6,6){\makebox(0,0){$\frac{P\times P}{G}$}}
 \put(6,2){\makebox(0,0){$ \frac{P}{G}$}}
 \put(1.9,5.5){\vector(0,-1){2.7}}
 \put(2.1,5.5){\vector(0,-1){2.7}}
 \put(5.9,5.5){\vector(0,-1){2.7}}
 \put(6.1,5.5){\vector(0,-1){2.7}}
 \put(4,6){\vector(1,0){0.8}}
 \put(3,2){\vector(1,0){2.2}}
 \put(2.5,5.5){\vector(4,-1){4.8}}
 \put(2.6,1.7){\vector(4,-1){4.7}}
 \put(6.4,2){\line(4,-1){5.9}}
 \put(6.3,1.9){\line(4,-1){5.9}}
 \put(7,6){\line(4,-1){5.5}}
 \put(6.9,5.9){\line(4,-1){5.5}}
 \put(4.7,4.4){\makebox(0,0){$\iota_2^* $}}
 \put(5,0.6){\makebox(0,0){$[\pi^*]$}}
 \put(9,4){\makebox(0,0){$\frac{P\times T^*_eG\times P}{G}$}}
 \put(9,0){\makebox(0,0){$P/G$}}
 \put(13,4){\makebox(0,0){$\frac{P\times P}{G}$}}
 \put(13,0){\makebox(0,0){$ \frac{P}{G}$}}
 \put(8.9,3.5){\vector(0,-1){2.7}}
 \put(9.1,3.5){\vector(0,-1){2.7}}
 \put(12.9,3.5){\vector(0,-1){2.7}}
 \put(13.1,3.5){\vector(0,-1){2.7}}
 \put(10.5,4){\vector(1,0){1.3}}
 \put(10.5,0){\vector(1,0){1.7}}
 \end{picture}\end{split}\end{gather}
which is the dual of the short exact sequence of (\ref{duzyVtrojkaG}) in sense of the dualization procedure described in Section~\ref{sect:prelims}. In particular the f\/irst groupoid of (\ref{C}) is the $\mathcal{V}\!\mathcal{B}$-groupoid dual to the tangent prolongation groupoid of the gauge groupoid $\frac{P\times P}{G}\tto P/G$.

\subsection{Symplectic leaves of duals of Atiyah sequences}\label{sec:sympleaves}

The results of Section \ref{sect:dAseq}, in particular Theorem~\ref{prop:lavessympl},
can be applied to the short exact sequence
 \begin{gather}\label{D}\begin{split}& \unitlength=5mm\begin{picture}(11,5.6)(-4.5,0)
 \put(-2,5){\makebox(0,0){$T^*\big(\frac{P\times P}{G}\big)$}}
 \put(5,5){\makebox(0,0){$\frac{T^*P\times T ^*P}{G}$}}
 \put(12,5){\makebox(0,0){$\frac{(P\times T^*_e G\times P)}{G}$}}
 \put(-2,0){\makebox(0,0){$\frac{P\times P}{G}$}}
 \put(5,0){\makebox(0,0){$\frac{P\times P}{G}$}}
 \put(12,0){\makebox(0,0){$\frac{P\times P}{G}$}}
 \put(5,4){\vector(0,-1){3}}
 \put(-2,4){\vector(0,-1){3}}
 \put(12,4){\vector(0,-1){3}}
 \put(7,5){\vector(1,0){2.6}}
 \put(6,0){\vector(1,0){4.6}}
 \put(-0.8,0){\vector(1,0){4.3}}
 \put(0,5){\vector(1,0){3}}
 \put(8.5,5.4){\makebox(0,0){$\iota_2^* $}}
 \put(8.5,0.7){\makebox(0,0){$\id$}}
\put(1.5,0.5){\makebox(0,0){$\id $}}
\put(1.5,5.4){\makebox(0,0){$\anch_2^* $}}
 \end{picture}\end{split}\end{gather}
of linear Poisson bundles; this is the dual of the Atiyah sequence of the $G$-principal bundle $(P\times P$, $\mu_2\colon P\times P\to \frac{P\times P}{G}, G)$. Note that (\ref{D}) is part of the $\mathcal{V}\!\mathcal{B}$-groupoid short exact sequence~(\ref{C}). In (\ref{C}) both $\iota_2^*$ and $\anch_2^*$ are Poisson maps, and are $\mathcal{V}\!\mathcal{B}$-groupoid morphisms. Summarizing the above, we can say that the structures of groupoid, vector bundle and Poisson manifold are consistently involved in the diagram (\ref{C}).

From the considerations in Section~\ref{sec:cotangauge}, especially of (\ref{C}), we see that the core of the $\mathcal{V}\!\mathcal{B}$-groupoid
\begin{gather*}
\begin{split}& \unitlength=5mm\begin{picture}(11,5.6)(-2,0)
 \put(1,5){\makebox(0,0){$T^*\big(\frac{P\times P}{G}\big)$}}
 \put(8,5){\makebox(0,0){$\frac{P\times P}{G}$}}
 \put(1,0){\makebox(0,0){$T^*P/G$}}
 \put(8,0){\makebox(0,0){$P/G$}}
 \put(1.2,4){\vector(0,-1){3}}
 \put(0.7,4){\vector(0,-1){3}}
 \put(8.2,4){\vector(0,-1){3}}
 \put(7.7,4){\vector(0,-1){3}}
 \put(3,5){\vector(1,0){3}}
 \put(2.4,0){\vector(1,0){3.7}}
 \put(2,2.4){\makebox(0,0){$\langle\Ss\rangle$}}
 \put(9.2,2.4){\makebox(0,0){$\langle \operatorname{pr}_2\rangle$}}
 \put(-0.1,2.4){\makebox(0,0){$\langle\Tt\rangle$}}
 \put(6.9,2.4){\makebox(0,0){$\langle \operatorname{pr}_1\rangle$}}
 \put(4.5,5.5){\makebox(0,0){$[\pi^*_2] $}}
 \put(4.5,0.5){\makebox(0,0){$[\pi^*] $}}
 \end{picture},\end{split}\end{gather*}
consists of all $\langle\varphi_p,0_p\rangle\in \frac{T^*P\times T^*P}{G}$ such that $\varphi_p\circ T\kappa_p(e)=0$, where $p\in P$. Thus, using~(\ref{Adual}), we have the following isomorphisms
\begin{gather*}
\core T^*\big(\tfrac{P\times P}{G}\big)\stackrel{\langle\Tt\rangle}{\longrightarrow}J^{-1}(0)/G\qquad {\rm and}\qquad T^*(P/G)\stackrel{\anch^*}{\longrightarrow}J^{-1}(0)/G
\end{gather*}
of vector bundles over $P/G$. In particular, the core of $T^*(\tfrac{P\times P}{G})$ is isomorphic to $T^*(P/G)$.

The vector bundle $T^*(P/G)\to P/G$ can be regarded as a totally intransitive groupoid on base $P/G$ and as such acts on $T^*P/G$ by
\begin{gather}\label{actionsympl}
T^*(P/G)*T^*P/G\ni(\rho_{\langle p\rangle}, \langle\varphi_p\rangle)\to \anch^*(\rho_{\langle p\rangle})+ \langle\varphi_p\rangle\in T^*P/G.
\end{gather}
 Recall here that $ \anch^*(\rho_{\langle p\rangle})$, $\langle\varphi_p\rangle\in [\pi^*]^{-1}(\langle p\rangle)$. The moment map for the action (\ref{actionsympl}) is $[\pi^*]\colon$ $T^*P/G\to P/G$, i.e., $[\pi^*](\langle\varphi_p\rangle)=\langle p\rangle\in P/G$ and
 \begin{gather*} T^*(P/G)*T^*P/G=\{(\rho,\langle\varphi\rangle)\in T^*(P/G)\times T^*P/G;\, \nu^*(\rho)=[\pi^*](\langle\varphi\rangle)\},\end{gather*}
 i.e., $(\rho,\langle\varphi\rangle)\in T^*(P/G)* T^*P/G$ if and only if $\langle q\rangle=\langle p\rangle$.

The action groupoid of (\ref{actionsympl}) is $T^*(P/G)\act_{[\pi^*]} T^*P/G\tto T^*P/G$ and according to~\cite[Proposition~11.2.3]{Mackenzie:GT}, valid for a general $\mathcal{V}\!\mathcal{B}$-groupoid, one has the short exact sequence of Lie groupoids
 \begin{gather}\label{sympl2}\begin{split}&\unitlength=5mm\begin{picture}(11,5.6)(-7,0) 
 \put(-3,5){\makebox(0,0){$T^*(P/G)\act_{[\pi^*]} T^*P/G$}}
 \put(5,5){\makebox(0,0){$T^*\big(\frac{P\times P}{G}\big)$}}
 \put(12,5){\makebox(0,0){$\frac{P\times P}{G}$}}
 \put(-2,0){\makebox(0,0){$T^*P/G$}}
 \put(5,0){\makebox(0,0){$T^*P/G$}}
 \put(12,0){\makebox(0,0){$P/G$}}
 \put(5,4){\vector(0,-1){3}}
 \put(5.3,4){\vector(0,-1){3}}
 \put(-2,4){\vector(0,-1){3}}
 \put(-1.7,4){\vector(0,-1){3}}
 \put(11.9,4){\vector(0,-1){3}}
 \put(12.2,4){\vector(0,-1){3}}
 \put(7,5){\vector(1,0){2.6}}
 \put(6.5,0){\vector(1,0){4.1}}
 \put(0,0){\vector(1,0){3}}
 \put(0.8,5){\vector(1,0){2}}
 \put(8.5,5.4){\makebox(0,0){$\ $}}
 \put(8.5,0.7){\makebox(0,0){$\ $}}
\put(1.5,0.5){\makebox(0,0){$\id $}}
\put(1.5,5.4){\makebox(0,0){$\ $}}
 \end{picture}\end{split}\end{gather}

There are other crucial statements which we collect in the following proposition:
\begin{Proposition}\label{prop:crucial19}\quad
 \begin{enumerate}\itemsep=0pt
\item[$(i)$] The groupoid $T^*\big(\frac{P\times P}{G}\big)\tto \frac{T^*P}{G}$ is the symplectic groupoid of the Poisson manifold $\frac{T^*P}{G}$.
\item[$(ii)$] The symplectic groupoid $T^*\big(\frac{P\times P}{G}\big)\tto \frac{T^*P}{G}$ is a subgroupoid as well as a symplectic leaf of the Poisson groupoid $\frac{T^*P\times T^*P}{G}\tto \frac{T^*P}{G}$.
\item[$(iii)$] Therefore the symplectic leaves of $\frac{T^*P}{G}$, defined as $J^{-1}(\O)/G$ are orbits of the standard action of the groupoid $T^*\big(\frac{P\times P}{G}\big)\tto \frac{T^*P}{G}$ on its base $\frac{T^*P}{G}$.
\item[$(iv)$] As a consequence of~\eqref{sympl2}, any symplectic leaf $J^{-1}(\O)/G$ is foliated by the orbits of the action of $T^*(P/G)\tto P/G$ on $T^*P/G$; these orbits are the fibres of the affine bundle $\iota^*\colon J^{-1}(\O)/G\to P\times _{\Ad^*_G}\O$.
\end{enumerate}
\end{Proposition}

Here (i) is a particular case of the theory introduced in~\cite{Weinstein:1987}.

The symplectic leaves $J_2^{-1}(\O)/G$ of $\frac{T^*P\times T^*P}{G}\tto \frac{T^*P}{G}$ may be included in a sequence of morphisms of f\/ibre bundles, shown in (\ref{E}), similarly to \eqref{part2}
\begin{gather}\label{E}\begin{split}&\unitlength=5mm\begin{picture}(11,5.6)(-4,0)
 \put(-2,5){\makebox(0,0){$T^*\big(\frac{P\times P}{G}\big)$}}
 \put(5,5){\makebox(0,0){$J_2^{-1}(\O)/G$}}
 \put(12,5){\makebox(0,0){$\frac{(P\times \O \times P)}{G}$}}
 \put(-2,0){\makebox(0,0){$\frac{P\times P}{G}$}}
 \put(5,0){\makebox(0,0){$\frac{P\times P}{G}$}}
 \put(12,0){\makebox(0,0){$\frac{P\times P}{G}$}}
 \put(5,4){\vector(0,-1){3}}
 \put(-2,4){\vector(0,-1){3}}
 \put(12,4){\vector(0,-1){3}}
 \put(7,5){\vector(1,0){2.6}}
 \put(6,0){\vector(1,0){4.6}}
 \put(-0.8,0){\vector(1,0){4.3}}
 \put(0,5){\vector(1,0){3}}
 \put(8.5,5.4){\makebox(0,0){$\iota_2^* $}}
 \put(8.5,0.7){\makebox(0,0){$\id$}}
\put(1.5,0.5){\makebox(0,0){$ \id $}}
\put(1.5,5.4){\makebox(0,0){$\anch_2^* $}}
 \end{picture}\end{split}\end{gather}
So, Proposition \ref{prop:lavessympl} is valid in this case as well as in the case of symplectic leaves $J^{-1}(\O)/G$, where $\O\in T^*_eG$. From (\ref{C}) we have
\begin{gather}\label{F}\begin{split}&\unitlength=5mm\begin{picture}(11,5.6)(-4,0)
 \put(-2,5){\makebox(0,0){$T^*\big(\frac{P\times P}{G}\big)$}}
 \put(5,5){\makebox(0,0){$J_2^{-1}(\O)/G$}}
 \put(12,5){\makebox(0,0){$\frac{(P\times \O \times P)}{G}$}}
 \put(-2,0){\makebox(0,0){$T^*P/G$}}
 \put(5,0){\makebox(0,0){$T^*P/G$}}
 \put(12,0){\makebox(0,0){$P/G$}}
 \put(5.2,4){\vector(0,-1){3}}
 \put(4.7,4){\vector(0,-1){3}}
 \put(-2.3,4){\vector(0,-1){3}}
 \put(-1.8,4){\vector(0,-1){3}}
 \put(12.2,4){\vector(0,-1){3}}
 \put(11.7,4){\vector(0,-1){3}}
 \put(7,5){\vector(1,0){2.6}}
 \put(6,0){\vector(1,0){4.6}}
 \put(-0.8,0){\vector(1,0){4.3}}
 \put(0,5){\vector(1,0){3}}
 \put(8.5,5.4){\makebox(0,0){$\iota_2^* $}}
 \put(8.5,0.7){\makebox(0,0){$[\pi^*]$}}
\put(1.5,0.5){\makebox(0,0){$ \id$}}
\put(1.5,5.4){\makebox(0,0){$\anch_2^* $}}
 \end{picture}\end{split}\end{gather}
Note that in diagram (\ref{F}) only $T^*\big(\frac{P\times P}{G}\big)\tto \frac{T^*P}{G}$ is a~groupoid. The target and source maps of $(T^*P\times T^*P)/G$ restricted to $J_2^{-1}(\O)/G$ def\/ine symplectic and anti-symplectic realizations of~$T^*P/G$, respectively. (In general these realizations are not full.) Thereby, we have the family of symplectic (anti-symplectic) realizations of $T^*P/G$, parametrized by the coadjoint orbits $\O\subset T^*_eG$.

\begin{Proposition}\label{prop:20} The action of the symplectic groupoid $T^*\big(\frac{P\times P}{G}\big)\tto \frac{T^*P}{G}$ on the symplectic realization $\Tt\colon J_2^{-1}(\O)/G\to \frac{T^*P}{G}$ is a symplectic action; that is, the graph of the action
\begin{gather}\label{map:20}
T^*\big(\tfrac{P\times P}{G}\big)\times_{T^*P/G} J_2^{-1}(\O)/G\to J_2^{-1}(\O)/G
\end{gather}
is Lagrangian and $\Tt\colon J_2^{-1}(\O)/G\to \frac{T^*P}{G}$ is a Poisson map which is equivariant with respect to this action and the natural action of $T^*\big(\tfrac{P\times P}{G}\big)$ on its base. Corresponding results with signs reversed hold for the action of $T^*\big(\frac{P\times P}{G}\big)\tto \frac{T^*P}{G}$ on $\Ss\colon J_2^{-1}(\O)/G\to \frac{T^*P}{G}$.
\end{Proposition}

\begin{proof} First consider the multiplication
\begin{gather}\label{mult:20}
\tfrac{T^*P\times T^*P}{G}\times_{T^*P/G} \tfrac{T^*P\times T^*P}{G}\to \tfrac{T^*P\times T^*P}{G}
\end{gather}
in the groupoid $\tfrac{T^*P\times T^*P}{G}\gpd T^*P/G$. This is the quotient over $G$ of the groupoid $T^*P\times T^*P\gpd T^*P$ (with the structure from (\ref{s})), which is a symplectic groupoid with respect to the form $d(\operatorname{pr}_1^*\gamma+\operatorname{pr}_2^*\gamma)$, where $\gamma$ is the canonical 1-form on $T^*P$; since $G$ preserves the symplectic structure, $\tfrac{T^*P\times T^*P}{G}\gpd T^*P/G$ is a Poisson groupoid. In particular, the graph of the multiplication~(\ref{mult:20}) is coisotropic in $\Bar{\gold}\times\gold\times\gold$ where we write $\gold = \tfrac{T^*P\times T^*P}{G}$ for brevity.

The map (\ref{map:20}) is a restriction of the multiplication (\ref{mult:20}). By (\ref{D}), $T^*\big(\tfrac{P\times P}{G}\big)$ is $(\iota_2^*)^{-1}(0)$ and $J_2^{-1}(\O)/G$ is $(\iota_2^*)^{-1}((P\times \O\times P)/G)$. Since $\iota_2^*$ is a morphism, it follows that the groupoid multiplication of an element of $T^*\big(\tfrac{P\times P}{G}\big)$ by an element of $J_2^{-1}(\O)/G$ is again an element of~$J_2^{-1}(\O)/G$.

Further, since $J_2^{-1}(\O)/G$ is a symplectic leaf in $\tfrac{T^*P\times T^*P}{G}$, and $T^*\big(\tfrac{P\times P}{G}\big)$ is a symplectic submanifold of $\gold$, the restriction of~(\ref{mult:20}) again has coisotropic graph, and by counting dimensions, is Lagrangian. Thus~(\ref{map:20}) is a symplectic action.
\end{proof}

\section[Applications in Hamiltonian mechanics of semidirect products]{Applications in Hamiltonian mechanics\\ of semidirect products}\label{sect:eacr}

In this section we investigate a few special cases of the results obtained in the two previous sections. We also indicate applications in the theory of Hamiltonian systems.

Let us begin with the simplest case when $P=G$. Then the dual Atiyah sequence (\ref{Adual}) reduces to the well-known isomorphism of Poisson manifolds $T^*G/G\cong T^*_eG$. The interest of the situation becomes clear if one considers the reciprocally dual short exact sequences of $\mathcal{V}\!\mathcal{B}$-groupoid presented in diagrams (\ref{duzyVtrojkaG}) and (\ref{C}).

Setting $P=G$ in (\ref{duzyVtrojkaG}) we obtain the following short exact sequence of $\mathcal{V}\!\mathcal{B}$-groupoids over the group(oid) $G\tto\{e\}$:
 \begin{gather}\label{ex:duzyVtrojkaG}\begin{split}&\unitlength=5mm\begin{picture}(11,8.6)(-6,0)
 \put(-5,8){\makebox(0,0){${TG}$}}
 \put(-5,4){\makebox(0,0){${ T_eG}$}}
 \put(-1,8){\makebox(0,0){$G$}}
 \put(-1,4){\makebox(0,0){$\{e\}$}}
 \put(-5.1,7.5){\vector(0,-1){2.7}}
 \put(-4.9,7.5){\vector(0,-1){2.7}}
 \put(-1.1,7.5){\vector(0,-1){2.7}}
 \put(-0.9,7.5){\vector(0,-1){2.7}}
 \put(-3,8){\vector(1,0){0.8}}
 \put(-4,4){\vector(1,0){2.2}}
 \put(-4.5,7.5){\vector(4,-1){4.4}}
 \put(-4.2,3.7){\vector(4,-1){5.2}}
 \put(-0,7.9){\line(4,-1){5}}
 \put(-0.1,7.8){\line(4,-1){5}}
 \put(-0.4,3.9){\line(4,-1){5.5}}
 \put(-0.5,3.8){\line(4,-1){5.5}}
 \put(-3,6.5){\makebox(0,0){$\iota_2$}}
 \put(-2,2.6){\makebox(0,0){$ \id$}}
 \put(2,6){\makebox(0,0){$\frac{TG\times TG}{G}$}}
 \put(2,2){\makebox(0,0){$T_eG$}}
 \put(6,6){\makebox(0,0){$G$}}
 \put(6,2){\makebox(0,0){$ \{e\}$}}
 \put(1.9,5.5){\vector(0,-1){2.7}}
 \put(2.1,5.5){\vector(0,-1){2.7}}
 \put(5.9,5.5){\vector(0,-1){2.7}}
 \put(6.1,5.5){\vector(0,-1){2.7}}
 \put(4,6){\vector(1,0){0.8}}
 \put(3,2){\vector(1,0){2.2}}
 \put(2.5,5.5){\vector(4,-1){4.8}}
 \put(2.6,1.7){\vector(4,-1){4.7}}
 \put(6.4,2){\line(4,-1){5.9}}
 \put(6.3,1.9){\line(4,-1){5.9}}
 \put(7,6){\line(4,-1){5.5}}
 \put(6.9,5.9){\line(4,-1){5.5}}
 \put(4.7,4.4){\makebox(0,0){$\anch_2 $}}
 \put(5,0.6){\makebox(0,0){$[\pi]$}}
 \put(9,4){\makebox(0,0){$TG$}}
 \put(9,0){\makebox(0,0){$\{e\}$}}
 \put(13,4){\makebox(0,0){$G$}}
 \put(13,0){\makebox(0,0){$ \{e\}$}}
 \put(8.9,3.5){\vector(0,-1){2.7}}
 \put(9.1,3.5){\vector(0,-1){2.7}}
 \put(12.9,3.5){\vector(0,-1){2.7}}
 \put(13.1,3.5){\vector(0,-1){2.7}}
 \put(10.5,4){\vector(1,0){1.3}}
 \put(10.5,0){\vector(1,0){1.7}}
 \end{picture}\end{split}\end{gather}

The central $\mathcal{V}\!\mathcal{B}$-groupoid in (\ref{ex:duzyVtrojkaG}) is the gauge groupoid of the principal bundle
\begin{gather*} (TG,G,\mu\colon \, TG\to TG/G\cong T_eG).\end{gather*} Note that the group $G$ acts on the tangent bundle $TG$ by the tangent lift of the right action of $G$ on itself. The f\/inal $\mathcal{V}\!\mathcal{B}$-groupoid of the short exact sequence (\ref{ex:duzyVtrojkaG}) is the tangent group $TG\tto\{e\}$ of the group $G$. The upper short exact sequence in (\ref{ex:duzyVtrojkaG}) is the Atiyah sequence of the principal $G$-bundle
\begin{gather*} \big(G\times G,G,\mu\colon \, G\times G\to \tfrac{G\times G}{G}\cong G\big) .
\end{gather*}

\begin{Remark}The initial $\mathcal{V}\!\mathcal{B}$-groupoid in (\ref{ex:duzyVtrojkaG}) has core zero and is a special case of the initial $\mathcal{V}\!\mathcal{B}$-groupoid in~(\ref{duzyVtrojkaG}). Note that there is no natural groupoid structure $T\gold\tto A\gold$ for a general Lie groupoid~$\gold$.
\end{Remark}

The dualization of (\ref{ex:duzyVtrojkaG}), according to (\ref{C}), leads to the dual sequence of $\mathcal{V}\!\mathcal{B}$-groupoids
 \begin{gather*}\label{ex:C}\begin{split}&\unitlength=5mm \begin{picture}(11,8.6)(-6,0)
 \put(-5,8){\makebox(0,0){$T^*G$}}
 \put(-5,4){\makebox(0,0){$T^*_eG$}}
 \put(-1,8){\makebox(0,0){$G$}}
 \put(-1,4){\makebox(0,0){$\{e\}$}}
 \put(-5.1,7.5){\vector(0,-1){2.7}}
 \put(-4.9,7.5){\vector(0,-1){2.7}}
 \put(-1.1,7.5){\vector(0,-1){2.7}}
 \put(-0.9,7.5){\vector(0,-1){2.7}}
 \put(-3,8){\vector(1,0){0.8}}
 \put(-4,4){\vector(1,0){2.2}}
 \put(-4.5,7.5){\vector(4,-1){4.4}}
 \put(-4.2,3.7){\vector(4,-1){5.2}}
 \put(-0,7.9){\line(4,-1){5}}
 \put(-0.1,7.8){\line(4,-1){5}}
 \put(-0.4,3.9){\line(4,-1){5.5}}
 \put(-0.5,3.8){\line(4,-1){5.5}}
 \put(-3,6.5){\makebox(0,0){$\anch_2^*$}}
 \put(-2,2.6){\makebox(0,0){$ \id$}}
 \put(2,6){\makebox(0,0){$\frac{T^*G\times T^*G}{G}$}}
 \put(2,2){\makebox(0,0){$T^*_eG$}}
 \put(6,6){\makebox(0,0){$G$}}
 \put(6,2){\makebox(0,0){$\{e\}$}}
 \put(1.9,5.5){\vector(0,-1){2.7}}
 \put(2.1,5.5){\vector(0,-1){2.7}}
 \put(5.9,5.5){\vector(0,-1){2.7}}
 \put(6.1,5.5){\vector(0,-1){2.7}}
 \put(4,6){\vector(1,0){0.8}}
 \put(3,2){\vector(1,0){2.2}}
 \put(2.5,5.5){\vector(4,-1){4.8}}
 \put(2.6,1.7){\vector(4,-1){4.7}}
 \put(6.4,2){\line(4,-1){5.9}}
 \put(6.3,1.9){\line(4,-1){5.9}}
 \put(7,6){\line(4,-1){5.5}}
 \put(6.9,5.9){\line(4,-1){5.5}}
 \put(4.7,4.4){\makebox(0,0){$\iota_2^*$}}
 \put(5,0.6){\makebox(0,0){$[\pi^*]$}}
 \put(9,4){\makebox(0,0){$ T^*G$}}
 \put(9,0){\makebox(0,0){$\{e\}$}}
 \put(13,4){\makebox(0,0){$G$}}
 \put(13,0){\makebox(0,0){$ \{e\}$}}
 \put(8.9,3.5){\vector(0,-1){2.7}}
 \put(9.1,3.5){\vector(0,-1){2.7}}
 \put(12.9,3.5){\vector(0,-1){2.7}}
 \put(13.1,3.5){\vector(0,-1){2.7}}
 \put(10.5,4){\vector(1,0){1.3}}
 \put(10.5,0){\vector(1,0){1.7}}
 \end{picture}\end{split}\end{gather*}
in which $\iota_2^*$ and $\anch_2^*$ are morphisms of Poisson manifolds.

We now consider the case in which the total space $P$ is a Lie group $H$ and $G=N\subset H$ is a~normal subgroup of $H$. This includes the case just treated. So, $K:=H/N$ is a group and one has the short exact sequence of Lie groups
\begin{gather*}
\{e\}\to N\stackrel{\iota}{\hookrightarrow} H\stackrel{\mu}{\to} K\to \{e\}.
\end{gather*}
We further assume that there is a global section $\sigma\colon K\to H$ of the principal bundle $\mu\colon H\to K$; we do not require $\sigma$ to be a morphism of groups. Without loss of generality we can assume that $\sigma(e_K)=e_H$, where $e_K$ and $e_H$ are the neutral elements of $K$ and $H$, respectively. Since $\sigma\colon K\to H$ is transversal to the f\/ibres of $\mu\colon H\to K$, we def\/ine an isomorphism of principal bundles $\Sigma\colon K\times N\to H$ (trivialization of $\mu\colon H\to K$) as follows
 \begin{gather}\label{exN:triv} \Sigma(k,u):=\sigma(k)\iota(u).\end{gather}
 From (\ref{exN:triv}) one obtains the decomposition
 \begin{gather}\label{exN:decomp} v_h=T(R_{\iota(u)}\circ \sigma)(k)\xi_k+T(L_{\sigma(k)}\circ\iota)(u)\nu_u\end{gather}
 of $v_h\in T_hH$ on $(\nu_u,\xi_k)\in T_kK \times T_uN$ which def\/ines the connection $\Gamma\colon TK\to TH$ by
 \begin{gather*} \Gamma(k)(\xi_k):=T(R_{\iota(u)}\circ\sigma)(k)\xi_k \end{gather*}
 and the connection form $\alpha\in\Gamma^\infty(TH,T_eN)$ by
 \begin{gather}\label{exN:form} \alpha(v_h)=T(\iota^{-1}\circ L_{h^{-1}})(h)\circ \left(\id_{T_hH}-T(R_{\iota(u)}\circ\sigma\circ \mu)\right)(h)v_h,\end{gather}
 where $\ u=\sigma(\mu(h))^{-1}h$.

 Let us note here that $T(R_{\iota(u)}\circ\sigma\circ \mu)(h)\colon T_hH\to T_hH$ is a projection of $T_hH$ on the horizontal subspace $T(R_{\iota(u)}\circ\sigma)(k)(T_kK)$ and $(\id_{T_hH}-T(R_{\iota(u)}\circ\sigma\circ \mu)(h))\colon T_hH\to T_hH$ is a projection of $T_hH$ on the vertical subspace $T(L_{\sigma(k)}\circ \iota)(T_uN)$ of the connection $\Gamma\colon TK\to TH$.

 Dualizing $\Gamma\colon TK\to TH$ we obtain $\Gamma^*\colon T^*H\to T^*K$ which for $\varphi_h\in T^*_hH$ is given by
 \begin{gather}\label{exN:Gdual} \Gamma^*(\varphi_h)=\varphi_h\circ T(R_h\circ \sigma)(k),\end{gather}
where $h=\sigma(k)\iota(u)$.

Note that the cotangent lift $T^*R_g\colon T^*_hH\to T^*_{hg}H$ of the right action $R_g\colon H\to H$ is given by
 \begin{gather}\label{exN:lift} T^*R_g(\varphi_h)=\varphi_h\circ(TR_g(h))^{-1}=\varphi_h\circ TR_{g^{-1}}(hg)\end{gather}
 and the corresponding momentum map $J_H\colon T^*H\to T^*_e H $ is given by{\samepage
 \begin{gather}\label{exN:momH}J_H(\varphi_h)=\varphi_h\circ TL_h(e)\end{gather}
 and satisf\/ies $ J_H\circ T^*R_g=\Ad^*_{g^{-1}}\circ J_H $ for $g\in H$.}

 Since for $u\in N$ one has $\Gamma^*\circ T^*R_{\iota(u)}=\Gamma^*$ the map (\ref{exN:Gdual}) def\/ines the map $[\Gamma^*]\colon T^*H/N\to T^*K$ which is not necessarily a Poisson map.

Using (\ref{exN:decomp}) we obtain the explicit formula of the map $T\Sigma\colon TK\times TN\to TH$
 \begin{gather}\label{exN:tangentriv} T\Sigma(\xi_k,\nu_u):=T(R_{\iota(u)}\circ \sigma)(k)\xi_k+T(L_{\sigma(k)}\circ \iota)(u)\nu_u,\end{gather}
where $\xi_k\in T_kK$ and $\nu_u\in T_uN$, tangent to the trivialization map $\Sigma\colon K\times N\to H$ def\/ined in~(\ref{exN:triv}). Since $T(L_{\sigma(k)}\circ\iota)(u)\nu_u\in ker T\mu(h)$ and~$\label{exN:prop1} \mu\circ R_{\iota(u)}\circ \sigma=\id_K$ we f\/ind that the inverse of~$T\Sigma$ is given by
\begin{gather*} (T\Sigma)^{-1}(v_h)=\big(T\mu(h)v_h,T\big(\iota^{-1}\circ L_{h^{-1}}\big)(h)\circ(\id_{T_hH}-T(R_{\iota(u)}\circ\sigma\circ \mu))(h)v_h\big),\end{gather*}
where $k=\mu(h)$, $\iota(u)=\sigma(\mu(h))^{-1}h$.

 The dif\/feomorphism $T^*\Sigma:=((T\Sigma)^{-1})^*\colon T^*K\times T^*N\to T^*H$ dual to $(T\Sigma)^{-1}$ takes the form
 \begin{gather*} T^*\Sigma(\theta_k,\chi_u)=\theta_k\circ T\mu(h)+\chi_u\circ T\big(\iota^{-1}\circ L_{h^{-1}}\big)(h)\circ\big(\id_{T_hH}-T(R_{\iota(u)}\circ\sigma\circ \mu)(h)\big)\end{gather*}
 on $(\theta_k,\chi_u)\in T_k^*K\times T_u^*N$. It factorizes $T^*H$ into the product of cotangent bundles $T^*K\times T^*N$ and allows us to simplify the form of the momentum map
\begin{gather*} J(\varphi_h)=\varphi_h\circ TL_h(e)\circ T\iota(e),\\
(J\circ T^*\Sigma)(\theta_k,\chi_u)=\chi_u\circ T(L_{\iota(u)}\circ\iota)(e).\end{gather*}

We take the pullback of the momentum map (\ref{exN:momH}) and the pullback of the action (\ref{exN:lift}) on $T^*K\times T^*N$. For $(J_H\circ T^*\Sigma)\colon T^*K\times T^*N\to T^*_eH$, where $h=\sigma(k)\iota(u)$, one has
\begin{gather*}
(J_H\circ T^*\Sigma)(\theta_k,\chi_u)=\theta_k\circ T(\mu\circ L_h)(e) \\
\hphantom{(J_H\circ T^*\Sigma)(\theta_k,\chi_u)=}{} +\chi_u\circ T(\iota^{-1}\circ L_{h^{-1}})(h)\circ\left(\id_{T_hH}-T(R_{\iota(u)}\circ\sigma\circ \mu)\right)(h)\circ TL_h(e).
\end{gather*}

 In order to obtain the explicit form of the action
\begin{gather*} T^*R_{(l,w)}:=(T^*\Sigma)^{-1}\circ T^*R_g\circ T^*\Sigma\colon \ T^*K\times T^*N\to T^*K\times T^*N\end{gather*}
 we take $g=\Sigma(l,w)$, where $(l,w)\in K\times N$, and use the product formula
\begin{gather} \label{exN:lift1}
 (k,u)(l,w):=\Sigma^{-1}(hg)= (kl,m(k,l)\varrho(l)(u)w ),
\end{gather}
 where the cocycle $m\colon K\times K\to N$ and the map $\varrho\colon K\to \Aut N$ are def\/ined by
 \begin{gather*} m(k,l):=\sigma(kl)^{-1}\sigma(k)\sigma(l)\qquad {\rm and}\qquad \varrho(l)(u):=\sigma(l)^{-1}u\sigma(l).\end{gather*}
 We simplify the problem assuming triviality of the cocycle $m\colon K\times K\to N$, i.e., we consider the case when $m\colon K\times K\to \{e\}$. Then $\varrho\colon K\to \Aut N$ is a group anti-homomorphism and $\sigma\colon K\to H$ is a group homomorphism. The group product (\ref{exN:lift1}) in this case assumes the following form
 \begin{gather}\label{exN:lift2} (k,u)(l,w)= (kl,\varrho(l)(u)w )=(kl,(R_{w}\circ \varrho(l))(u)),\end{gather}
and $H$ is the semidirect product $K\ltimes_\rho N$.
 The product (\ref{exN:lift2}) leads to the formula{\samepage
 \begin{gather}\label{exN:lift3} T^*R_{(l,w)}(\theta_k,\chi_u)=\big(\theta_k\circ TR_{l^{-1}}(kl),\chi_u\circ T\big[(R_{\iota(w)}\circ\varrho(l)\circ\iota)^{-1}\big](u)\big)\end{gather}
 for the action $T^*R_{(l,w)}\colon T_k^*K\times T^*_uN\to T^*_{kl}K\times T^*_{(R_{\iota(w)}\circ\varrho(l)\circ\iota)(u)}N$.}

 Taking in (\ref{exN:tangentriv}) $k=e$ and $u=e$ we obtain the isomorphism
 \begin{gather}\label{exN:iso} T\Sigma_e(\xi_e,\nu_e):=T\sigma(e)\xi_e+T\iota(e)\nu_e\end{gather}
 of $T_eK\times T_eN$ with $T_eH$ and, thus the decomposition
 \begin{gather*}
 T_eH=T\sigma(e)(T_eK)\oplus T\iota(e)(T_eN).\end{gather*}
 Superposing the dual $(T\Sigma_e)^*\colon T_e^*H\to T_e^*K\times T_e^*N$ of (\ref{exN:iso}) with $J_H\circ T^*\Sigma\colon T^*K\times T^*N\to T^*_eH$ we f\/ind that the factorized momentum map
\begin{gather*} J_\Sigma:=(T\Sigma_e)^*\circ J_H\circ T^*\Sigma\colon \ T^*K\times T^*N\to T^*_eK\times T^*_eN\end{gather*}
is given by
\begin{gather}
J_\Sigma(\theta_k,\nu_u)=(\theta_k\circ TL_k(e),\chi_u\circ T(L_{\iota(u)}\circ\iota)(e))\nonumber\\
\hphantom{J_\Sigma(\theta_k,\nu_u)}{} =(J_K(\theta_k),(J\circ T^*\Sigma)(\theta_k,\chi_u))=(J_K(\theta_k),J_N(\chi_u)),\label{exN:decomp2}
\end{gather}
where $J_K\colon T^*K\to T^*_eK$ and $J_N\colon T^*N\to T^*_eN$ are the momentum maps for $K$ and $N$, respectively.
 Combining (\ref{exN:lift3}) and (\ref{exN:decomp2}) we obtain the following equivariance property
 \begin{gather}\label{exN:morphism0} J_\Sigma\circ T^*R_{(l,w)}=\big(\Ad^*_{l^{-1}}\times \Ad^*_{w^{-1}}\circ\big(T\varrho\big(l^{-1}\big)(e)\big)^*\big)\circ J_\Sigma,\end{gather}
for $J_\Sigma$, where
\begin{gather*} \label{exN:morphism}K\ltimes_\varrho N\ni(l,w) \mapsto\big(\Ad^*_{l^{-1}}\times \Ad^*_{w^{-1}}\circ\big(T\varrho\big(l^{-1}\big)(e)\big)^*\big)\in \Aut(T^*_eK\times T^*_eN)\end{gather*}
 is an anti-homomorphism of $K\ltimes_\varrho N\cong H$ in $\Aut(T^*_eK\times T^*_eN)$.

 Note that $J_\Sigma$ is the momentum map related to the symplectic form $d((T^*\Sigma)_*\gamma_H)$ where $(T^*\Sigma)_*\gamma_H$ is the pullback of the canonical $1$-form $\gamma_H$ of $T^*H$ by $T^*\Sigma\colon T^*K\times T^*N\to T^*H$. The subsequent proposition expresses $(T^*\Sigma)_*\gamma_H$ in terms of the canonical form $\gamma_K$ of $T^*K$ and the connection form $\alpha$ def\/ined in~(\ref{exN:form}).
 \begin{Proposition} One has the following equality
 \begin{gather} ((T^*\Sigma)_*\gamma_H)(\theta_k,\chi_u)=(\operatorname{pr}_K^*)_*\gamma_K(\theta_k)\nonumber\\
 \hphantom{((T^*\Sigma)_*\gamma_H)(\theta_k,\chi_u)=}{} +\chi_u\circ TL_{\iota(u)}(e)\circ(\Sigma\circ[(\pi_K^*\circ \operatorname{pr}_K^*)\times (\pi_N^*\circ \operatorname{pr}_N^*)])_*\alpha, \label{exN:pulback}\end{gather}
 where $\pi^*_K\colon T^*K\to K$ and $\pi^*_N\colon T^*N\to N$ are the projections on the bundle bases, and $\operatorname{pr}_K^*\colon$ $T^*K\times T^*N\to T^*K$ and $\operatorname{pr}_N^*\colon T^*K\times T^*N\to T^*N$ are the canonical projections.
 \end{Proposition}
 \begin{proof}
 For $X_{(\theta_k,\chi_u)}\in T_{(\theta_k,\chi_u)}(T^*K\times T^*N)$ we have
\begin{gather}
 \langle((T^*\Sigma)_*\gamma_H)(\theta_K,\chi_u),X_{(\theta_k,\chi_u)}\rangle
 = \langle(\gamma_H((T^*\Sigma)(\theta_k,\chi_u)),T(T^*\Sigma)(\theta_k,\chi_u)X_{(\theta_k,\chi_u)}\rangle\nonumber\\
 \qquad{} =\langle((T^*\Sigma)(\theta_k,\chi_u),
 T\pi^*((T^*\Sigma)(\theta_k,\chi_u))T(T^*\Sigma)(\theta_k,\chi_u)X_{(\theta_k,\chi_u)}\rangle\nonumber\\
 \qquad{} =\langle((T^*\Sigma)(\theta_k,\chi_u),
 T(\pi^*\circ T^*\Sigma)(\theta_k,\chi_u))X_{(\theta_k,\chi_u)}\rangle\nonumber\\
\qquad{} =\langle\theta_k\circ T\mu(h)+\chi_u\circ TL_{\sigma(k)^{-1}}(h)\nonumber\\
 \qquad\quad{} \circ(\id_{T_hH}-T(R_{\iota(u)}\circ\sigma\circ\mu)(h)),
 T(\pi^*\circ T^*\Sigma)(\theta_k,\chi_u))X_{(\theta_k,\chi_u)}\rangle\nonumber\\
 \qquad{} =\langle\theta_k,T(\mu\circ \pi^*\circ T^*\Sigma)(\theta_k,\chi_u)X_{(\theta_k,\chi_u)}\rangle\nonumber\\
 \qquad\quad{} +
 \langle \chi_u, TL_{\iota(u)}(e)\circ \alpha\circ T(\pi^*\circ T^*\Sigma)(\theta_k,\chi_u)X_{(\theta_k,\chi_u)}\rangle,\label{exN:pulback1}
\end{gather}
where for the last equality we used (\ref{exN:form}). Now we observe that
\begin{gather}\label{exN:1}
\pi^*\circ T^*\Sigma=\Sigma\circ((\pi^*_K\circ \operatorname{pr}_K^*)\times (\pi_N^*\circ \operatorname{pr}_N^*))
\end{gather}
and
\begin{gather}\label{exN:2}
\nu\circ \pi^*\circ T^*\Sigma=\pi^*_K\circ \operatorname{pr}^*_K.
\end{gather}
Substituting (\ref{exN:1}) and (\ref{exN:2}) into (\ref{exN:pulback1}) we obtain
\begin{gather*}
\langle((T^*\Sigma)_*\gamma_H)(\theta_K,\chi_u),X_{(\theta_k,\chi_u)}\rangle
=\langle\theta_k,T(\pi^*_K\circ \operatorname{pr}^*_K)(\theta_k,\chi_u)X_{(\theta_k,\chi_u)}\rangle\nonumber\\
\qquad\quad{} +\langle \chi_u, TL_{\iota(u)}(e)\circ \alpha\circ T(\Sigma\circ((\pi^*_K\circ \operatorname{pr}_K^*)\times (\pi_N^*\circ \operatorname{pr}_N^*)))(\theta_k,\chi_u)X_{(\theta_k,\chi_u)}\rangle\nonumber\\
\qquad{} =\langle(\operatorname{pr}_K^*)_*\gamma_K,X_{(\theta_k,\chi_u)}\rangle+\langle \chi_u\circ TL_{\iota(u)}(e)\circ \alpha\circ T(\Sigma\circ((\pi^*_K\circ \operatorname{pr}_K^*)\\
\qquad\quad{} \times (\pi_N^*\circ \operatorname{pr}_N^*)))_*(\theta_k,\chi_u),X_{(\theta_k,\chi_u)}\rangle,
\end{gather*}
which gives (\ref{exN:pulback}).
\end{proof}

The f\/irst term in (\ref{exN:pulback}) is the pullback of $\gamma_K$ and the second one,
called the ``magnetic term'', is the pullback of the connection form $\alpha$ (superposed with $\chi_u\circ TL_{\iota(u)}(e)\in T^*_e N)$ on the product $T^*K\times T^*N$. Let us note that $(\operatorname{pr}^*_K)_*\gamma_K$ as a section of $T^*(T^*K\times T^*N)$ is linear on the f\/ibres of $T^*K$ and constant on the second factor of $T^*K\times T^*N$. Similarly, the magnetic term is linear on the f\/ibres of $T^*N$, constant on the f\/ibres of $T^*K$, but not constant on the base $K\times N$ of $T^*K\times T^*N$.

Using the dif\/feomorphism $T^*\Sigma\colon T^*K\times T^*N\stackrel{\sim}{\to} T^*H$ we can write the dual Atiyah sequence~(\ref{Adual}) for $P=H=K\ltimes_\rho N$ and $G=N$ as follows
\begin{gather*} 0\longrightarrow T^*K\stackrel{\anch^*}{\longrightarrow}T^*K\times T^*_eN \stackrel{\iota^*}{\longrightarrow} K\times T^*_eN\longrightarrow 0, \end{gather*}
where $\anch^*(\theta_k)=(\theta_k, 0)$ and $\iota^*(\theta_k, \chi_l)=(k,\chi_l)$. Hence the symplectic leaves of $T^*K\times T^*_eN$ are
\begin{gather*} T^*K\times \O\cong (J\circ T^*\Sigma)^{-1}(\O)/N={\iota^*}^{-1}(K\times\O)/N,\end{gather*}
where $\O\subset T^*_eN$ are coadjoint orbits of $N$. One obtains the symplectic structure
of the leaf $T^*K\times \O$ by the reduction to ${\iota^*}^{-1}(K\times\O)\subset T^*K\times T^*N$ of $d((T^*\Sigma)_*\gamma_H)$ presented in~(\ref{exN:pulback}).

Since $J_\Sigma\colon T^*K\times T^*N\to T^*_eK\times T^*_eN\cong T^*_eH$
is a Poisson map, where $T^*_eH$ is equipped with the linear Poisson structure, any function $f\in C^\infty(T^*_eK\times T^*_eN)$ def\/ines a Hamiltonian $f\circ J_\Sigma$ on $T^*K\times T^*N$. According to (\ref{exN:morphism0}), if $f$ is invariant with respect to the action
\begin{gather*} \{e\}\times N\ni(e,w)\mapsto\big(\id\times \Ad^{-1}_{w^{-1}}\big)\in \Aut(T^*K\times T^*_eN).\end{gather*}
Then $f\circ J_\Sigma\in C^\infty(T^*K\times T^*N/N)\cong C^\infty(T^*K\times T^*_eN)$ def\/ines a Hamiltonian system on the Poisson manifold $T^*K\times T^*_eN$. This is always so if $N$ is an abelian group.

In this case the coadjoint representation $\Ad^*\colon N\to \Aut T^*_eN$ of $N$ is trivial, i.e., $\Ad^*=\id$. So, its orbits are one-element subsets $\{a\}\subset T^*_eN$ of $T^*_eN$. It follows from Theorem \ref{prop:lavessympl} that $[\Gamma^*]\colon T^*H/N\to T^*K$ def\/ines dif\/feomorphisms $[\Gamma^*]\colon J^{-1}(a)/N\to T^*K$, see also (\ref{pisigma}), of the symplectic leaves $J^{-1}(a)/N$, $a\in T^*_eN$, with the cotangent bundle $T^*K$.

 From point (ii) of Theorem \ref{prop:lavessympl} we see that $J^{-1}(a)/N\to H\times _{\Ad^*_N}\{a\}\cong K$ is an af\/f\/ine bundle over $K$. The dif\/feomorphism $[\Gamma^*]\colon J^{-1}(a)/N\stackrel{\sim}{\to} T^*K$ preserves the f\/ibre bundle structures, i.e., it covers the identity map of the base $K$. Let us note here that
\begin{gather*} J^{-1}(a)/N\cong (J\circ T^*\Sigma)^{-1}((0,a))/N\cong T^*K\times J_N^{-1}(a)/N\cong T^*K\times \{0\}\cong T^*K . \end{gather*}
 The symplectic form $\omega_a$ of $(J\circ T^*\Sigma)^{-1}((0,a))$ is obtained as the reduction of $d(T^*\Sigma)_*\gamma_H$. So, from (\ref{exN:morphism0}) and the dif\/feomorphism $J^{-1}(0)/N\cong T^*K$ we f\/ind that
\begin{gather*}
 \omega_a=d(\gamma_K+(\pi_K^*)_*A),
\end{gather*}
where $A$ is a $1$-form on $K$ def\/ined by the connection $1$-form $\alpha$ and $a\in T^*_eN$, and $(\pi_K^*)_*A$ is the pullback of $A$ on $T^*K$ by $\pi^*_K\colon T^*K\to K$.

As an example one can present the case of the heavy top, where $K={\rm SO}(3)$, $N=\mathbb{R}^3$ and $\rho\colon {\rm SO}(3)\to \Aut\mathbb{R}^3$ is the usual action of ${\rm SO}(3)$ on $\mathbb{R}^3$. For detailed description of this case see for example~\cite{ratiu2}. For the description of $n$-dimensional tops in the framework of semi-direct product Hamiltonian mechanics we refer the reader to~\cite{ReymanSTS}. One can f\/ind many other examples of application of semidirect product Hamiltonian mechanics in \cite{ratiu1, holm, ratiu2} which concern also inf\/inite dimensional physical systems.

 The short exact sequence of $\mathcal{V}\!\mathcal{B}$-groupoids (\ref{C}) in the case when $P=H\cong K\ltimes_\varrho N$ and $G=N$ assumes the following form
 \begin{gather}\label{exN:C}\begin{split}&\unitlength=5mm \begin{picture}(11,8.6)(-8.5,0)
 \put(-6,8){\makebox(0,0){\fontsize{9pt}{1pt}{$T^*(K\times N\times K)$}}}
		\put(-6,4){\makebox(0,0){\fontsize{9pt}{1pt}{$T^*K\times T^*_eN$}}}
 \put(0,8){\makebox(0,0){\fontsize{9pt}{1pt}{${K\times N\times K}$}}}
 \put(-1,4){\makebox(0,0){$K$}}
 \put(-5.1,7.5){\vector(0,-1){2.7}}
 \put(-4.9,7.5){\vector(0,-1){2.7}}
 \put(-1.1,7.5){\vector(0,-1){2.7}}
 \put(-0.9,7.5){\vector(0,-1){2.7}}
 \put(-3.4,8){\vector(1,0){1.3}}
 \put(-4,4){\vector(1,0){2.2}}
 \put(-4.3,7.3){\vector(4,-1){4.4}}
 \put(-4.2,3.7){\vector(4,-1){5.2}}
 \put(0.5,7.5){\line(4,-1){4.7}}
 \put(0.9,7.6){\line(4,-1){4.7}}
 \put(-0.4,3.9){\line(4,-1){5.5}}
 \put(-0.5,3.8){\line(4,-1){5.5}}
 \put(-3,6.5){\makebox(0,0){$\anch_2^* $}}
 \put(-2,2.6){\makebox(0,0){$ \id$}}
 \put(1.4,6){\makebox(0,0){\footnotesize{$T^*K\!\times\!\frac{T^*N\!\times\! T^*N}{N}\!\times\! T^*K$}}}
 \put(1,2){\makebox(0,0){\footnotesize{$T^*K\times T^*_eN$}}}
 \put(7.5,6){\makebox(0,0){\footnotesize{${K\times N\times K}$}}}
 \put(6,2){\makebox(0,0){$K$}}
 \put(1.9,5.5){\vector(0,-1){2.7}}
 \put(2.1,5.5){\vector(0,-1){2.7}}
 \put(5.9,5.5){\vector(0,-1){2.7}}
 \put(6.1,5.5){\vector(0,-1){2.7}}
 \put(4.5,6){\vector(1,0){1.1}}
 \put(3,2){\vector(1,0){2.2}}
 \put(2.5,5.5){\vector(4,-1){4.8}}
 \put(2.6,1.7){\vector(4,-1){4.7}}
 \put(6.4,2){\line(4,-1){5.9}}
 \put(6.3,1.9){\line(4,-1){5.9}}
 \put(7.5,5.6){\line(4,-1){5}}
 \put(7.8,5.7){\line(4,-1){5}}
 \put(4.7,4.4){\makebox(0,0){$\iota_2^* $}}
 \put(5,0.6){\makebox(0,0){$\ $}}
 \put(8.5,4){\makebox(0,0){\footnotesize{$K\!\times\! T^*N\!\times\! K$}}}
 \put(9,0){\makebox(0,0){$K$}}
 \put(14,4){\makebox(0,0){\footnotesize{${K\times N\times K}$}}}
 \put(13,0){\makebox(0,0){$K$}}
 \put(8.9,3.5){\vector(0,-1){2.7}}
 \put(9.1,3.5){\vector(0,-1){2.7}}
 \put(12.9,3.5){\vector(0,-1){2.7}}
 \put(13.1,3.5){\vector(0,-1){2.7}}
 \put(10.5,4){\vector(1,0){1.3}}
 \put(10.5,0){\vector(1,0){1.7}}
 \end{picture}\end{split}\end{gather}
where we used the following dif\/feomorphisms
\begin{gather*}
K\ltimes_\varrho N/N\cong K,\\
\frac{(K\ltimes_\varrho N)\times (K\ltimes_\varrho N) }{N}\cong K\times N\times K,\\
T^*(K\ltimes_\varrho N)/N\cong T^*K\times T^*_eN,\\
\frac{T^*((K\ltimes_\varrho N)\times (K\ltimes_\varrho N))}{N}\cong T^*K\times\frac{T^*N\times T^*N}{N}\times T^*K,\\
\frac{(K\ltimes_\varrho N)\times T^*_eN\times (K\ltimes_\varrho N))}{N}\cong K\times T^*N \times K.
\end{gather*}

Let us note that we can consider (\ref{exN:C}) as a product of the short exact sequence of $\mathcal{V}\!\mathcal{B}$-groupoids (\ref{C}), taken for $P=K$ and $G=\{e\}$, with the short exact sequence of $\mathcal{V}\!\mathcal{B}$-groupoids (\ref{C}) taken for $P=N$ and $G=N$. But we have to stress here that the Poisson structures in (\ref{exN:C}) are not the products of the respective Poisson structures and they depend on the anti-homomorphism $\varrho\colon K\to \Aut N$.

To end, we mention that one can consider the Poisson manifolds which are included in the upper dual Atiyah sequence of (\ref{exN:C}) as the phase spaces of some composed systems related to the ones presented in the lower dual Atiyah sequence of~(\ref{exN:C}).

\subsection*{Acknowledgements}
We extend our best thanks to the referees, who caught many slips and obscurities.


\pdfbookmark[1]{References}{ref}
\LastPageEnding


\begin{thebibliography}{99}
\footnotesize\itemsep=0pt

\bibitem{Wei}
Cannas~da Silva A., Weinstein A., Geometric models for noncommutative algebras,
 \textit{Berkeley Mathematics Lecture Notes}, Vol.~10, Amer. Math. Soc.,
 Providence, RI, Berkeley Center for Pure and Applied Mathematics, Berkeley,
 CA, 1999.

\bibitem{CDW}
Coste A., Dazord P., Weinstein A., Groupo\"\i des symplectiques, in
 Publications du {D}\'epartement de {M}ath\'ematiques. {N}ouvelle {S}\'erie.
 {A}, {V}ol.~2, \textit{Publ. D\'ep. Math. Nouvelle S\'er. A}, Vol.~87,
 University Claude-Bernard, Lyon, 1987, 1--62.

\bibitem{ratiu1}
Gay-Balmaz F., Ratiu T.S., The geometric structure of complex f\/luids,
 \href{https://doi.org/10.1016/j.aam.2008.06.002}{\textit{Adv. in Appl. Math.}} \textbf{42} (2009), 176--275,
 \href{https://arxiv.org/abs/0903.4294}{arXiv:0903.4294}.

\bibitem{holm}
Holm D.D., Marsden J.E., Ratiu T.S., The {E}uler--{P}oincar\'e equations and
 semidirect products with applications to continuum theories, \href{https://doi.org/10.1006/aima.1998.1721}{\textit{Adv.
 Math.}} \textbf{137} (1998), 1--81, \href{https://arxiv.org/abs/chao-dyn/9801015}{chao-dyn/9801015}.

\bibitem{Mackenzie:GT}
Mackenzie K.C.H., General theory of {L}ie groupoids and {L}ie algebroids,
 \href{https://doi.org/10.1017/CBO9781107325883}{\textit{London Mathematical Society Lecture Note Series}}, Vol.~213, Cambridge
 University Press, Cambridge, 2005.

\bibitem{MW}
Marsden J., Weinstein A., Reduction of symplectic manifolds with symmetry,
 \href{https://doi.org/10.1016/0034-4877(74)90021-4}{\textit{Rep. Math. Phys.}} \textbf{5} (1974), 121--130.

\bibitem{ratiu2}
Marsden J.E., Ratiu T.S., Weinstein A., Semidirect products and reduction in
 mechanics, \href{https://doi.org/10.2307/1999527}{\textit{Trans. Amer. Math. Soc.}} \textbf{281} (1984), 147--177.

\bibitem{Pradines:1988}
Pradines J., Remarque sur le groupo\"{\i}de cotangent de {W}einstein--{D}azord,
 \textit{C.~R.~Acad. Sci. Paris S\'er.~I Math.} \textbf{306} (1988), 557--560.

\bibitem{ReymanSTS}
Reyman A.G., Semenov-Tian-Shansky M.A., Group-theoretical methods in the theory
 of f\/inite-dimensional integrable systems, in Dynamical Systems~VII,
 \href{https://doi.org/10.1007/978-3-662-06796-3_7}{\textit{Encyclopaedia of Mathematical Sciences}}, Vol.~16, Springer, Berlin~--
 Heidelberg, 1994, 116--225.

\bibitem{Stern}
Sternberg S., Minimal coupling and the symplectic mechanics of a classical
 particle in the presence of a~{Y}ang--{M}ills f\/ield, \textit{Proc. Nat. Acad.
 Sci. USA} \textbf{74} (1977), 5253--5254.

\bibitem{Weinstein:1977}
Weinstein A., A universal phase space for particles in {Y}ang--{M}ills f\/ields,
 \href{https://doi.org/10.1007/BF00400169}{\textit{Lett. Math. Phys.}} \textbf{2} (1977), 417--420.

\bibitem{Weinstein:1987}
Weinstein A., Symplectic groupoids and {P}oisson manifolds, \href{https://doi.org/10.1090/S0273-0979-1987-15473-5}{\textit{Bull. Amer.
 Math. Soc. (N.S.)}} \textbf{16} (1987), 101--104.

\end{thebibliography}
\end{document}